\documentclass[useAMS,usenatbib]{mn2e}

\usepackage{graphicx}
\usepackage{multirow}
\usepackage{pifont}
\usepackage{mathrsfs}
\usepackage{amsmath}
\usepackage{url}
\usepackage{mathrsfs}
\usepackage[ansinew]{inputenc}
\usepackage{bm}
\usepackage{bbm}
\usepackage{bbold}
\usepackage{subfigure}
\usepackage{graphics}
\usepackage{graphicx}
\usepackage{hyperref}
\usepackage{xfrac}
\usepackage{array}
\usepackage[all]{xy}
\usepackage{epic}
\usepackage{latexsym}
\usepackage{enumerate}
\usepackage{appendix}
\usepackage{color}


\title[Large-step neural network] 
{Large-step neural network for learning the symplectic evolution from partitioned data}
\author[Xin Li, Jian Li, Zhihong Jeff Xia and Nikolaos Georgakarakos]
{Xin Li$^{1}$, Jian Li$^{2,3}$\thanks{E-mail: ljian@nju.edu.cn}, Zhihong Jeff Xia$^{4}$ and Nikolaos Georgakarakos$^{5,6}$\\
$^1$ Department of Statistics and Data Science, Southern University of Science and Technology of China. No 1088, Xueyuan Rd.,  Xili,\\~~~  Nanshan District, Shenzhen, Guangdong, 518055, PR China,\\
$^2$School of Astronomy and Space Science, Nanjing University, 163 Xianlin Avenue, Nanjing 210023, PR China,\\
$^3$Key Laboratory of Modern Astronomy and Astrophysics in Ministry of Education, Nanjing University, Nanjing 210023, PR China\\
$^4$Department of Mathematics, Northwestern University, 2033 Sheridan Road, Evanston, IL  60208, USA\\
$^5$New York University Abu Dhabi, PO Box 129188 Abu Dhabi, United Arab Emirates\\
$^6$ Center for Astro, Particle and Planetary Physics (CAP3), New York University Abu Dhabi, PO Box 129188 Abu Dhabi, United Arab Emirates}
\begin{document}

\date{Accepted 1988 December 15. Received 1988 December 14; in original form 1988 October 11}

\pagerange{\pageref{firstpage}--\pageref{lastpage}} \pubyear{2002}

\maketitle

\label{firstpage}

\begin{abstract}

In this study, we focus on learning Hamiltonian systems, which involves predicting the coordinate ($\bm q$) and momentum ($\bm p$) variables generated by a symplectic mapping. Based on \citet{CT21}, the symplectic mapping is represented by a generating function. To extend the prediction time period, we develop a new learning scheme by splitting the time series ($\bm q_i$, $\bm p_i$) into several partitions. We then train a large-step neural network (LSNN) to approximate the generating function between the first partition (i.e. the initial condition) and each one of the remaining partitions. This partition approach makes our LSNN effectively suppress the accumulative error when predicting the system evolution. Then we train the LSNN to learn the motions of the 2:3 resonant Kuiper belt objects for a long time period of 25000 yr. The results show that there are two significant improvements over the neural network constructed in our previous work \citep{li22}: (1) the conservation of the Jacobi integral, and (2) the highly accurate predictions of the orbital evolution. Overall, we propose that the designed LSNN has the potential to considerably improve predictions of the long-term evolution of more general Hamiltonian systems.

 \end{abstract}

\begin{keywords}
celestial mechanics -- Kuiper belt: general -- planets and satellites: dynamical evolution and stability -- methods: miscellaneous
\end{keywords}

\section{Introduction}

Neural networks were initially created to recognize patterns by identifying the appropriate relationship between covariates (i.e. inputs) and variates (i.e. outputs), operating as a black-box \citep{McCu1943, Rosen1958}. These artificial neural networks (ANNs) are inspired by the human brain and have the capacity to solve complex problems. ANNs  comprise of an interconnected network that can learn from vast amounts of experiential data by adjusting the weights on their connections. This approach has proven highly effective when being applied to a broad range of pattern recognition problems in science and industry. 

Time series prediction has become an intensive field of academic research. Traditional approaches to this problem have constructed parametric models based on domain expertise, e.g. auto-regressive (AR) \citep{BJ1976}, exponential smoothing \citep{Winters1960, G1985} and structural time series models \citep{Harvey1990}. However, with the development of machine learning, a new approach has emerged that can learn the temporal dynamics in a data-driven manner \citep{Ahmed2010}. The rapid advancement in computing power has a significant impact on time series forecasting and machine learning techniques now play a crucial role in this respect.

Differential equations have been widely used for studying most physical models. In order to predict time series, modern neural networks, or more precisely, deep ANNs (i.e. with more than one hidden layer), are trained to represent continuous dynamical systems \citep{E17}. A key challenge in deep machine learning is to learn the solutions of the differential equations as time series; and to go a step further, a more advanced task is to learn the differential equations themselves, represented by a class of functions.

In the realm of astronomy and astrophysics, machine learning has already been widely applied to exoplanet discovery \citep{scha19, arms21}, galaxy classification \citep{zhan19, baqu21, vavi21}, and dark matter research \citep{agar18, luci19, petu21}. However, for the classical $n$-body problem in celestial mechanics, relevant studies have only begun very recently. \citet{Grey2019} studied the Hamiltonian systems. They designed ``Hamiltonian neural networks (HNN)'' to learn the laws of physics directly from data and the conservation of the energy-like quantities over a long timescale. The trained HNN was used to predict the coordinate and momentum variables of future orbits. However, their method did not work well for the 3-body problem, which is actually non-integrable. In a subsequent study, \citet{Breen2020} claimed to have solved the general 3-body problem by training a deep ANN. For the planar case, their deep ANN can provide solutions as accurate as those obtained from numerical integrations and it costs much less computation time. But we notice that their best-performing ANN was only trained for trajectories with quite a short evolution time of $T=3.9$ time units\footnote{In the model of a triple system consisting of an inner binary and a third body, the dimensionless unit of time is $2\pi$, which corresponds to one orbital period of the inner binary.}. For more references, readers are referred to the latest review of applications of machine learning to asteroid dynamics by \citet{carr22}.

The fundamental difficulty in learning the motion of the 3-body problem is due to chaos, as there are parts of the phase space of the 3-body problem where slight changes in the initial conditions can result in significantly different long-term trajectories \citep{Poin1892}. In the framework of the planar circular restricted 3-body problem (PCR3BP), we studied the behaviours of objects in the 2:3 mean motion resonance with Neptune, which can be considered as an intermediate case between integrable and chaotic dynamical systems \citep{li22}. By learning the solutions of the differential equations in terms of orbital elements, our designed ANN can predict the trajectories of the 2:3 resonators over an entire resonant period of up to 25000 yr. Initially, we tried to train the ANN using the coordinate and momentum variables, but we found that the predicted values can not guarantee the conservation of the Jacobi integral. Therefore, we opted to use orbital elements as our training data, because this way we actually assumed that the system is a weakly perturbed 2-body problem. With this trick, we may predict the long-term trajectory along which the Jacobi integral is almost conserved. However, the weakness of employing the orbital elements is that the trick of using the 2-body approximation cannot be applied to many Hamiltonian systems, such as non-hierarchical triple systems or physical systems other than the $n$-body problem \citep{li09}. If our training data set consists of the coordinate and momentum variables instead of orbital elements, we can extend our machine learning technique to predict the long-term evolution of more general Hamiltonian systems with high accuracy.

Very recently, \citet{CT21} introduced a novel approach to learning and predicting non-linear time series generated by latent symplectic mappings. They studied Hamiltonian systems whose solution flows give rise to such symplectic mappings. They predicted the future trajectory by representing the symplectic mapping via a generating function, which is approximated by the so-called generating function neural network (GFNN). \citet{CT21} have compared their method with several other results that have been successful in learning Hamiltonian systems, such as HNN \citep{Grey2019, Bertalan21}, SRNN \citep{Chen2020}, SympNets \citep{Jin2020},  \citet{Lutter2019}, \citet{Toth2020}, \citet{Zhong2020}, \citet{Wu2020}, and \citet{Xiong2021}. However, all of these networks, except for SympNets, are designed to learn quantities that produce the Hamiltonian vector field. The results of the comparison show that GFNN outperforms the other methods on all occasions. In addition, \citet{CT21} proved that the prediction error of their method grows at most linearly with time. This is a remarkable contribution because errors of other methods can grow exponentially. According to these merits, it seems the GFNN is the state-of-the-art technique.

We now consider whether we could extend this efficient technique of using the generating function to learn the symplectic evolution from the coordinate and momentum variables. Based on GFNN, we are about to make a major adjustment to the structure of learning data: for the trajectories in the learning data set, we divide each trajectory into several partitions and then deduce the generating function between the first and one of the remaining partition elements. By doing so, we can train the ANN to learn a class of generating functions for the entire trajectory, i.e., the time series evolution. Since each partition element, although at different time windows, is equally (symplectically) mapped from the first one, the accuracy of our prediction should be kept at a similar level but not increase very fast over time. As we will show later, for the same learning data set from the PCR3BP provided by \citet{CT21}, the overall loss of our ANN we can achieve is smaller than that of the GFNN. In particular, our partition approach allows the developed ANN to handle time series evolution with large time steps. Therefore, given a fixed number of time points, we can predict future orbits over extended periods of time with high accuracy, characterised by the preservation of the Jacobi integral in the PCR3BP. The long-term prediction serves as the main advantage of this work, while we would not try to improve the GFNN itself.

The rest of this paper is organised as follows: In Section 2, we provide the mathematical background on linear Hamiltonian systems, including the symplectic mapping and generating function, as well as the design of our ANN for predicting future time series. In Section 3, for the same dynamical model in the framework of the PCR3BP, we compare the performance of our ANN with the GFNN proposed by \citet{CT21}. In Section 4, we apply our ANN to the evolution of the non-resonant and resonant Kuiper belt objects (KBOs), we demonstrate that the predicted orbits can have small errors even over quite a long time interval and that the Jacobi integral is also well conserved. The conclusions and discussion are presented in Section 5.

\section{Methods}

\subsection{Hamiltonian system}

Let us begin by reviewing some basic mathematical background of Hamiltonian systems \citep{MHO09}. We define $gl(m, \mathbb{R})$ as the set of all $m \times m$ non-singular matrices with entries in the real number field $\mathbb{R}$. Suppose $\bm{z}$ is a coordinate vector in $\mathbb{R}^{2n}$, $\mathbb{I}$ is a time interval in $\mathbb{R}$, and the matrix $K: \mathbb{I} \rightarrow gl(2n,\mathbb{R})$ is continuous and symmetric. Then, a Hamiltonian system can be described by $2n$ ordinary differential equations: 
 \begin{equation}\label{H}
 \dot{\bm{z}} = J \frac{\partial H}{\partial \bm{z}} = J K(t)\bm{z}=A(t)\bm{z},
 \end{equation} 
 where $H=H(\bm{z},t)$ is called the Hamiltonian of the system, and $J$ is the $2n \times 2n$ Poisson matrix:
 \begin{equation}\label{J}
 J = \begin{bmatrix}0_n & -I_n \\ I_n & 0_n \end{bmatrix},
 \end{equation}
with $I_n$ being the $n\times n$ identity matrix and $0_n$ being the $n\times n$ zero matrix.
 
Let $t_0 \in \mathbb{I}$ be the initial time point. According to the theory of differential equations, for equation (\ref{H}), if the Lipschitz condition in the variable $\bm{z}$ is satisfied, there exists a unique solution $\bm z(t,t_0,\bm{z}_0)$ for $t \in \mathbb{I}$ given the initial condition $\bm{z}_0=\bm{z}(t_0) \in \mathbb{R}^{2n}$. The time evolution $\phi$ from $\bm{z}(t_i)$ to $\bm{z}(t_{i+1})$ is said to be symplectic if and only if the Jacobian matrix $\phi^{\prime}=\partial \bm{z}(t_{i+1})/\partial \bm{z}(t_i)$ satisfies the condition
 \begin{equation}\label{phi}
 (\phi^{\prime})J(\phi^{\prime})^T = J.
 \end{equation}
As for the Hamiltonian equation (\ref{H}), this condition is naturally satisfied.

In the current work, only the latent evolution of a Hamiltonian system is considered. Let 
\begin{equation}\label{pq}
\bm{z}(t)=(\bm{q}(t), \bm{p}(t)),
 \end{equation}
where $\bm{q}$ is the coordinate, $\bm{p}$ is the momentum, and these two variables are said to be conjugate. Then, the dynamical system described by equation (\ref{H}) can be expressed in the following form:  
 \begin{equation}\label{Hpq}
\dot{\bm{p}}(t)=-\frac{\partial H}{\partial \bm{q}(t)},~~~\dot{\bm{q}}(t)=\frac{\partial H}{\partial \bm{p}(t)}.
 \end{equation}
Assuming that each trajectory $(\bm{q}(t), \bm{p}(t))$ is sampled at discrete time points $t_i$ with a time step $h~(>0)$, we have 
\begin{equation}\label{piqi}
\bm{q}_i = \bm{q}(t_i),~~~\bm{p}_i = \bm{p}(t_i)~~~~(t_i=ih).
 \end{equation}
If we consider the one step mapping
\begin{equation}\label{phi}
\phi: (\bm{q}(t_i), \bm{p}(t_i)) \rightarrow (\bm{q}(t_{i+1}), \bm{p}(t_{i+1})), ~~~\forall t_i,
 \end{equation}
since $(\bm{q}(t), \bm{p}(t))$ represents the symplectic flow, the mapping $\phi$ is symplectic.

In our previous study \citep{li22}, the solution of the restricted 3-body problem was learned by a multi-layer ANN. We originally tried to train the ANN with the coordinate and momentum variables (i.e. $\bm{q}$ and $\bm{p}$). Although the ANN's prediction could be very close to the ground truth, we observed significant variations in the Jacobi integral throughout the orbital evolution. To overcome this issue, we now adopt a new approach inspired by \citet{CT21}, whereby we construct an ANN to learn a class of symplectic mappings $\phi$ rather than directly training it on $\bm{q}$ and $\bm{p}$. The symplectic mappings $\phi$ allow us to derive the time series $(\bm{q}(t_i), \bm{p}(t_i))$, preserving the system's symplectic structure.We think this technique may potentially help us improve the conservation of the Jacobi integral. Of course, we also need to control the errors of the predicted $\bm{q}$ and $\bm{p}$ on a small scale. Indeed, we can  regard $\phi$ in equation (\ref{phi}) as a canonical transformation and then we will introduce the classical method of the ``generating function'' in what follows.

\subsection{Symplectic mapping and generating function}

Here we consider the multidimensional vectors [$\bm q$,$\bm p$], where $\bm q,\bm p \in \mathbb{R}^n$. Corresponding to the 2$n$-dimensional Hamiltonian system, $\bm q$ is the coordinate variable and $\bm p$ is the conjugate momentum variable. Then, the standard symplectic form is defined to be 
\begin{equation}\label{Omega}
\Omega = d\bm q \wedge d\bm p,
 \end{equation}
where the operator $\wedge$ is called the exterior product or wedge product. If $\bm Q$ and $\bm P$ are the new coordinate and momentum respectively, we illustrate substitutions of the form
\begin{equation}\label{pqPQ}
\bm Q = \bm Q(\bm q, \bm p), \bm P = \bm P(\bm q, \bm p).
 \end{equation}

The change of variables in equation (\ref{pqPQ}) is volume-preserving if and only if
\begin{equation}\label{change0}
d\bm q \wedge d\bm p = d\bm Q \wedge d\bm P,
 \end{equation}
which actually indicates that the transformation $(\bm q, \bm p) \rightarrow (\bm Q, \bm P)$ is symplectic. From equation (\ref{change0}), we can easily have
\begin{equation}\label{change2}
d\bm p\wedge d\bm q+d\bm Q\wedge d\bm P = 0.
 \end{equation}
This is sufficient to  make sure that the form $\bm pd\bm q + \bm Qd\bm P$ is exact, i.e. we can find a function $F$ that satisfies 
\begin{equation}\label{dF}
dF(\bm q, \bm P) = \bm pd\bm q+\bm Qd\bm P.
 \end{equation}
Since the differential form of $F$ can be written as 
\begin{equation}\label{dF2}
dF=\frac{\partial F}{\partial \bm q}d\bm q+\frac{\partial F}{\partial \bm P}d\bm P,
 \end{equation}
we can obtain the change of variables from $(\bm{q},\bm{p})$ to $(\bm{Q},\bm{P})$, by
\begin{equation}\label{change}
\bm p=\frac{\partial F}{\partial \bm q}(\bm{q},\bm{P}),\quad \bm Q=\frac{\partial F}{\partial \bm P}(\bm{q},\bm{P}).
 \end{equation}

According to the implicit function theorem, when the Hessian matrix $\frac{\partial^2 F}{\partial \bm q\partial \bm P}$ is non-singular and $\bm q$, $\bm p$ are not arbitrarily far apart $\bm P$, $\bm Q$, equation (\ref{change}) can provide a local solution for $\bm P$ and $\bm Q$ as functions of $\bm q$ and $\bm p$. Therefore, equation (\ref{change}) gives a straightforward way to construct the symplectic transformation and the function $F$ is usually called the generating function. For more details, we refer the reader to Chapter 6 in \citet{MHO09}.

\subsection{Large-step neural network (LSNN)}

The GFNN developed by \citet{CT21} is a neural network designed to learn the generating function in order to obtain a symplectic mapping from one time point to the next. This enables the system to evolve continuously over time, as the mapping can be repeated step by step. More specifically, the symplectic mapping for a single time step can be described as:  
$$(\bm q_0,\bm p_0) \rightarrow (\bm Q_1,\bm P_1).$$
If we denote 
$$\tilde{\bm q}_1 =\bm Q_1,\quad \tilde{\bm p}_1 =\bm P_1,$$ 
then the next mapping is
$$(\tilde{\bm q}_1,\tilde{\bm p}_1) \rightarrow (\bm Q_2,\bm P_2).$$
This process could go on until 
$$(\tilde{\bm q}_{i-1},\tilde{\bm p}_{i-1}) \rightarrow (\bm Q_i, \bm P_i),$$ 
where the subscript $i~(=3,\dots,N)$ corresponds to the $i$-th time point $t_i$ in equation (\ref{phi}). We note that the error of the above prediction may exhibit a linear increase as $N$ becomes larger, i.e. over a longer period of time.

\begin{figure}
 \hspace{0cm}
  \centering
  \includegraphics[width=8.2cm]{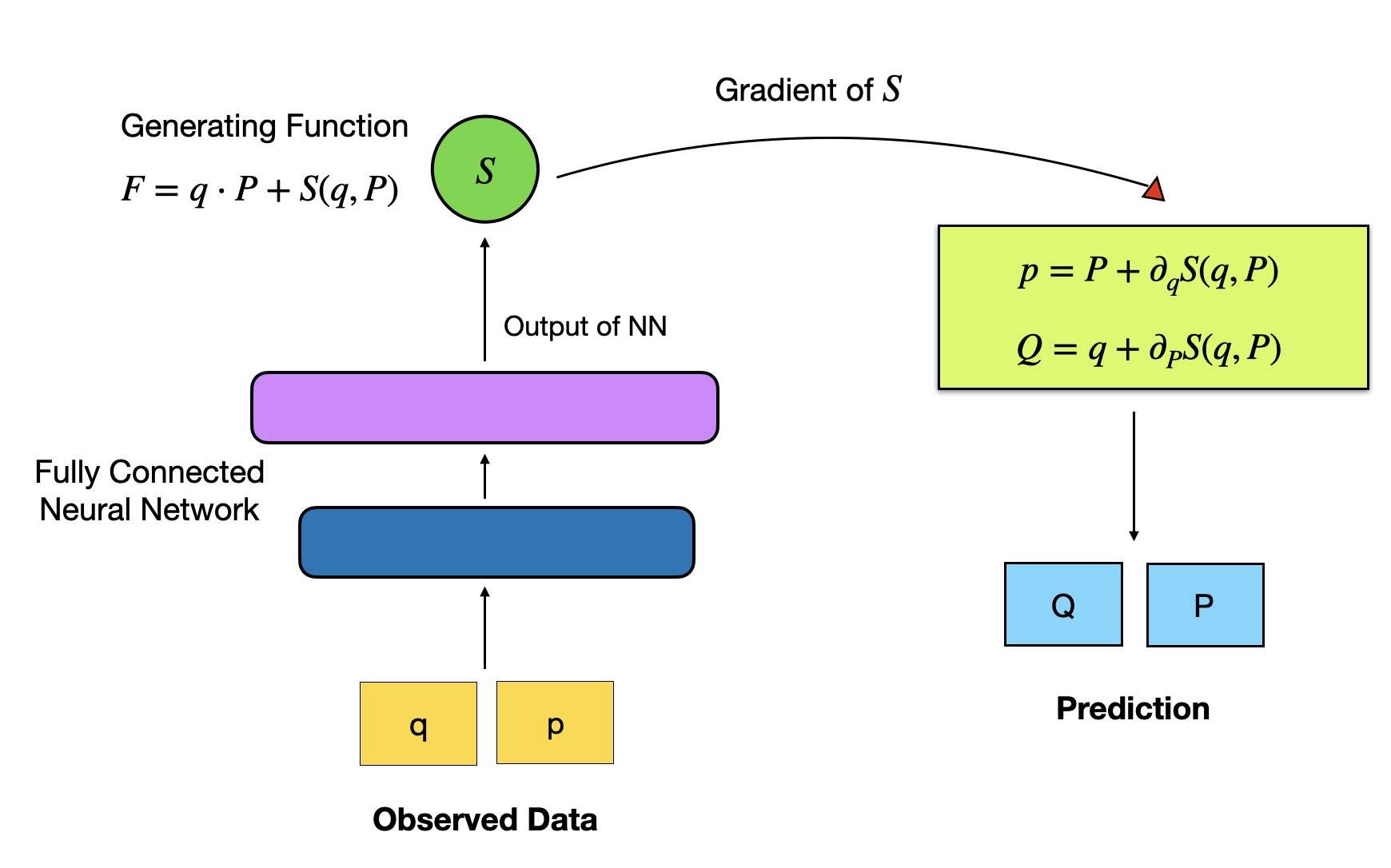}
\caption{The structure of LSNN. The observed data consisting of the coordinate $\bm q$ and momentum $\bm p$ is the input. The neural network contains 6 fully connected layers and learns to conserve a quantity which approximates the modified generating function $S$. Then we use $S$ to predict the future orbit points ($\bm Q$, $\bm P$) by the symplectic mapping.}
  \label{Schedule}
\end{figure}

\begin{figure}
 \hspace{0cm}
  \centering
  \includegraphics[width=6cm]{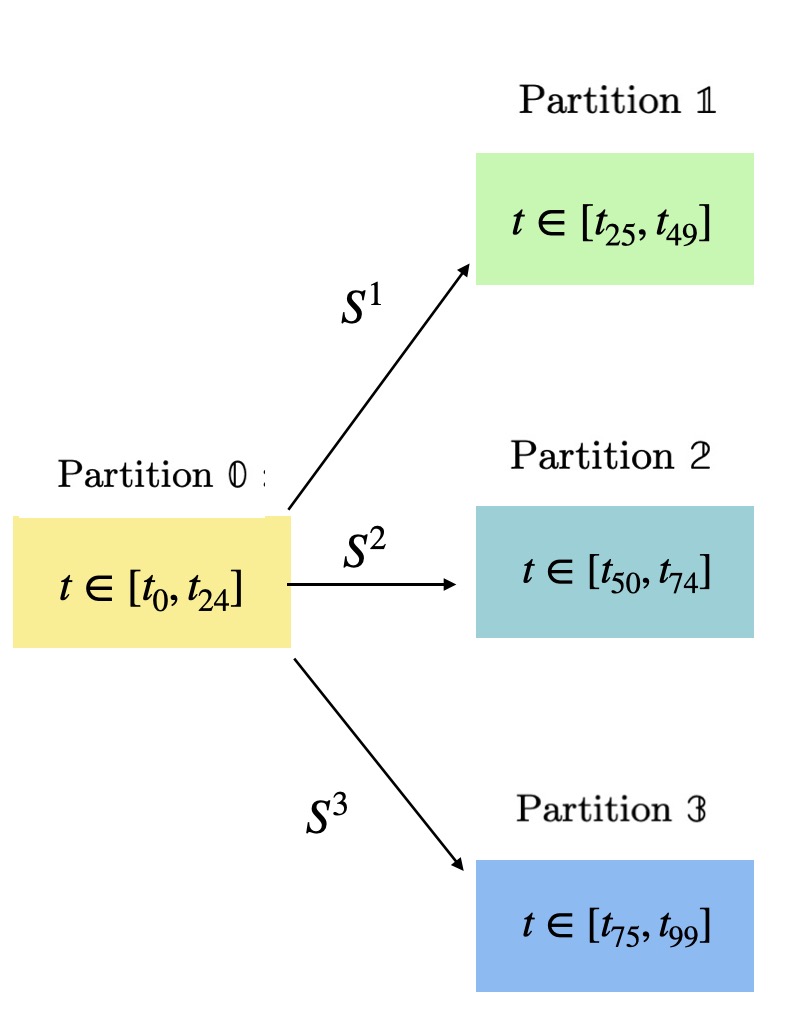}
\caption{We split the time series $(\bm q_0,\bm p_0),\dots,(\bm q_{99},\bm p_{99})$ at respective time points $t\in[t_0,\dots,t_{99}]$ into 4 partitions. Partition $\mathbb 0$ is set to be the initial condition, which is used to make the prediction for each of the remaining partitions ($\mathbb 1$, $\mathbb 2$ and $\mathbb 3$) separately. The prediction is achieved by learning the modified generating function $S$ as we illustrated in Fig. \ref{Schedule}. Since each $S^j~(j=\mathbb 1, \mathbb 2, \mathbb 3)$ represents the symplectic transformation between Partition $\mathbb 0$ and Partition $j$, the prediction errors may not accumulate from Partition $\mathbb 1$ at the early time to Partition $\mathbb 3$ at the late time.}
  \label{partition}
\end{figure}

Suppose that we have the training set containing a time series in the interval $[t_0, t_{99}]$ (i.e. $N=99$) with a fixed time step $h$. Differently from \citet{CT21}, we now consider 4 partitions of the time series $({\bm q}_i,{\bm p}_i)$: 
$$\text{Partition } \mathbb 0: i=0,\cdots,24,~~~\mbox{i.e.}~t \in[t_0, t_{24}],$$ 
$$\text{Partition } \mathbb 1: i=25,\cdots,49,~~~\mbox{i.e.}~t \in[t_{25}, t_{49}],$$ 
$$\text{Partition } \mathbb 2: i=50,\cdots,74,~~~\mbox{i.e.}~t \in[t_{50}, t_{74}],$$ 
$$\text{Partition } \mathbb 3: i=75,\cdots,99,~~~\mbox{i.e.}~t \in[t_{75}, t_{99}],$$
where Partition $\mathbb 0$ is set to be the initial condition. Our new approach is to train a neural network to learn the generating function $F$ from partitioned data. Once $F$ is approximated, we can use it to predict the time series. This process is visualized in Figs. \ref{Schedule} and \ref{partition}.  

To start, we establish mappings between the elements in Partition $\mathbb 0$ and Partition $\mathbb 1$ using the generating function 
$$F^1(\bm q,\bm P) = \bm q\cdot \bm P +S^1(\bm q,\bm P),$$
where $\bm q= \bm q_i$, $\bm P=\bm p_{i+25},~(i=0,...,24)$, and the modified generating function $S^1$ is the same for any $i$. 
When the Hessian matrix $\frac{\partial^2 F^1}{\partial \bm q \partial\bm P}$ is non-singular, our neural network can learn $S^{1}$ with respect to the loss function
\begin{equation*}
\begin{split}
\mathcal L_{S^1} &= \frac{1}{25\cdot M}\sum_{k=1}^M\sum_{i=0}^{24}\\
&(||\partial_{\bm p_{i+25,k}} S^1(\bm q_{i,k},\bm p_{i+25,k})-(\bm q_{i+25,k}-\bm q_{i,k})||_2^2\\
&+||\partial_{\bm q_{i,k}} S^1(\bm q_{i,k},\bm p_{i+25,k})- (\bm p_{i,k}-\bm p_{i+25,k})||_2^2),
\end{split}
\end{equation*}
where the operator $||\cdot||_2$ is the Euclidian 2-norm and $M$ is the total number of orbits.

When $S^1$ is learned, Partition $\mathbb 1$ can be predicted by
\begin{equation}
\label{predictionP1}
\bm p=\bm P + \partial_{\bm q} S^{1}(\bm q,\bm P),
\end{equation}
\begin{equation}
\label{predictionQ1}
\bm Q=\bm q +  \partial_{\bm P} S^{1}(\bm q, \bm P).
\end{equation} 
These expressions are equivalent to
\begin{equation}
\label{predictionP2}
\bm p_i=\tilde{\bm p}_{i+25} + \partial_{\bm q_i} S^{1}(\bm q_i, \tilde{\bm p}_{i+25}),
\end{equation}
\begin{equation}
\label{predictionQ2}
{\bm q}_{i+25}=\bm q_i +  \partial_{\tilde{\bm p}_{i+25}} S^{1}(\bm q_i, \tilde{\bm p}_{i+25}),
\end{equation}
where $i=0,...,24$. Accordingly we can obtain the following symplectic mappings
\begin{equation}\label{P1}
(\bm q_0,\bm p_0) \rightarrow (\tilde{\bm q}_{25},\tilde{\bm p}_{25}) ,\dots, (\bm q_{24},\bm p_{24}) \rightarrow (\tilde{\bm q}_{49},\tilde{\bm p}_{49}).
 \end{equation}

In order to predict Partition $\mathbb 1$ using the method described above, we need to know the momenta $\bm p_{25},\cdots,\bm p_{49}$, which actually have no ground truth values. Hence, we first train an additional fully connected neural network to predict the approximate momenta $\bm p'_{25},\cdots,\bm p'_{49}$ by the supervised learning like we did in \citet{li22}. Taking the individual $\bm p'_{25},\cdots,\bm p'_{49}$ as the initial conditions, we then use Newton's iteration to solve each of the following nonlinear equations (i.e. equation (\ref{predictionP2})):
\begin{eqnarray}\label{newton}
~~~~~~~~~~\tilde{\bm p}_{25} &=& \bm p_0 - \partial_{\bm q_0} S^{1}(\bm q_0, \tilde{\bm p}_{25}),\nonumber\\
&\vdots&\nonumber\\
\tilde{\bm p}_{49} &=& \bm p_{24} - \partial_{\bm q_{24}} S^{1}(\bm q_{24}, \tilde{\bm p}_{49}).
\end{eqnarray}
By this procedure we could get the predicted momenta $\tilde{\bm p}_{25},\dots,\tilde{\bm p}_{49}$. Subsequently, the predicted coordinates $\tilde{\bm q}_{25},\dots,\tilde{\bm q}_{49}$ can be obtained from equation (\ref{predictionQ2}).

It is worth noting that the initial values $\bm p'_{25},\cdots,\bm p'_{49}$ used in Newton's iteration are derived through the supervised learning and they are independent of the step size $h$ between two successive time points $t_i$ and $t_{i+1}$. This implies that our neural network may deal with the cases with large values of $h$. However, in \citet{CT21}, the authors use the initial value ${\bm p}_i$ for Newton's iteration in order to compute the next time series point $\tilde{\bm p}_{i+1}$ from the equation
\begin{equation}
\tilde{\bm p}_{i+1} = \bm p_i - \partial_{\bm q_i} S(\bm q_i, \tilde{\bm p}_{i+1})
\end{equation}
and thus a small time step of $h$ (i.e. 0.1) between $t_i$ and $t_{i+1}$ is obviously required.

In the exact same way, Partition $\mathbb 2$ and Partition $\mathbb 3$ can also be predicted from Partition $\mathbb 0$ by learning the individual generating functions: 
$$F^2(\bm{q},\bm{P}) = \bm q\cdot \bm P +S^2(\bm{q},\bm{P}),$$
where $\bm q= \bm q_i$, $\bm P=\bm p_{i+50}$ $(i=0,...,24)$ and
$$F^3(\bm{q},\bm{P}) = \bm q\cdot \bm P +S^3(\bm{q},\bm{P}),$$
where $\bm q= \bm q_i$, $\bm P=\bm p_{i+75}$ $(i=0,...,24)$. 

Then we can obtain the remaining two groups of symplectic mappings
\begin{equation}\label{P2}
(\bm q_0,\bm p_0) \rightarrow (\tilde{\bm q}_{50},\tilde{\bm p}_{50}),\dots, (\bm q_{24},\bm p_{24}) \rightarrow (\tilde{\bm q}_{74},\tilde{\bm p}_{74}),
 \end{equation}
and
\begin{equation}\label{P3}
(\bm q_0,\bm p_0) \rightarrow (\tilde{\bm q}_{75},\tilde{\bm p}_{75}),\dots, (\bm q_{24},\bm p_{24}) \rightarrow (\tilde{\bm q}_{99},\tilde{\bm p}_{99}).
 \end{equation}

To summarise, when considering Partitions $\mathbb 1$, $\mathbb 2$ and $\mathbb 3$, if ANNs are designed to approximate the modified generating functions $S^1$, $S^2$ and $S^3$ with errors $\leq \epsilon$ in the first derivative, then the losses between the predicted sequence $\{(\tilde{\bm q}_i, \tilde{\bm p}_i)\}$ and the ground truth sequence $\{(\bm q_i, \bm p_i)\}$ satisfy the following inequalities:
$$||\bm p_i - \tilde{\bm p}_i||_2  \leq \alpha \cdot\epsilon,$$ and 
\begin{equation}\label{global_loss}
||\bm q_i - \tilde{\bm q}_i||_2 \leq \alpha \cdot\epsilon, 
\end{equation}
where $i\in[25, 99]$ and $\alpha$ is some constant.

The learning scheme designed above will be called the large-step neural network (LSNN). The first motivation for constructing the LSNN is to suppress the accumulative error at a low level. As it will be demonstrated later, this can be achieved by utilising the symplectic mappings (\ref{P1}), (\ref{P2}) and (\ref{P3}) for Partitions $\mathbb 1$, $\mathbb 2$ and $\mathbb 3$ respectively. We further note that the LSNN has the theoretical capacity to predict the orbit evolution over an extended time of $\ge t_{100}$, i.e. for the subsequent Partitions $\mathbb 4$, $\mathbb 5$ and so on. However, accomplishing this may not be a simple task, as we will demonstrate in Section 3 when learning a specific data set. As such, we focuses on the case of 100 time points from $t_0$ to $t_{99}$. More importantly, even with a fixed number of 100 time points, the LSNN enables accurate predictions over a long timescale because it allows for a large time step $h$ between two successive time points. Although the learning technique is based on GFNN, the major advantage of our LSNN lies in its data structure feature. This advantage is expected to have important practical applications in long-term prediction.

\subsection{Prediction losses}

The sequence $({\bm q}_i,{\bm p}_i)$ under consideration consists of 100 time points, indexed from $i=0$ to 99, and has been divided into 4 partitions. Therefore, for each of the mappings (\ref{P1}), (\ref{P2}), and (\ref{P3}), the LSNN model  predicts a time series with a length of 25, e.g. $i=25$-$49$ for $\text{Partition } \mathbb 1$. In addition, since in the subsequent dynamical models we only consider the planar case, either ${\bm q}_i$ or ${\bm p}_i$ will be assumed to have two components.

Then, for each of Partitions $\mathbb 1$, $\mathbb 2$ and $\mathbb 3$, we can define two $25\times 2$ matrices as:
$$\bm Q_{sim}^j(25\times 2)=\{\bm q_{25\cdot j},\bm q_{25\cdot j+1}...,\bm q_{25\cdot j+24}\},$$ 
$$\bm P_{sim}^j(25\times 2)=\{\bm p_{25\cdot j},\bm p_{25\cdot j+1}...,\bm p_{25\cdot j+24}\},$$
where the subscript ``sim'' denotes the ground truth data from numerical simulations and $j$ indicates the number of the partition. Similarly, we can also define the matrices predicted by the LSNN as: 
$$\bm Q_{LSNN}^j(25\times 2)=\{\tilde{\bm q}_{25\cdot j},\tilde{\bm q}_{25\cdot j+1}...,\tilde{\bm q}_{25\cdot j+24}\},$$
$$\quad \bm P_{LSNN}^j(25\times 2)=\{\tilde{\bm p}_{25\cdot j},\tilde{\bm p}_{25\cdot j+1}...,\tilde{\bm p}_{25\cdot j+24}\}.$$

To evaluate the performance of the LSNN for a specific partition, we compute the average loss between the predicted data (i.e. $\bm Q_{LSNN}^j$ and $\bm P_{LSNN}^j$) and the corresponding ground truth data (i.e. $\bm Q_{sim}^j$ and $\bm P_{sim}^j$). If we have two $25\times 2$ matrices $A$ and $B$, the distance between them can be defined as:
$$\mathcal D(A,B)=\frac{1}{25}\sum_{k=1}^{25} \frac{\sqrt{\sum_{l=1}^2(A_{kl}-B_{kl})^2}}{2}.$$ 
Accordingly, for a single trajectory, we define the \textit{partitioned} single losses on each partition element as follows:
\begin{eqnarray}\label{loss1}
\text{$loss(\bm Q^j)$} &=& \mathcal D(\bm Q^j_{LSNN}(25\times 2) ,\bm Q^j_{sim}(25\times 2)),\nonumber\\
\text{$loss(\bm P^j)$} &=& \mathcal D(\bm P^j_{LSNN}(25\times 2) ,\bm P^j_{sim}(25\times 2)).
\end{eqnarray}
When we combine all three partitions for $i=25$-$99$, the single losses can be defined by
\begin{eqnarray}\label{loss3}
\text{$loss (\bm Q)$} &=& \frac{\sum_{j=1}^3 loss(\bm Q^j)}{3},\nonumber\\
\text{$loss (\bm P)$} &=& \frac{\sum_{j=1}^3 loss({\bm P^j})}{3}.
\end{eqnarray}

Note that equation (\ref{loss3}) provides only the single losses for a particular orbit. To calculate the global losses for multiple orbits, we take the arithmetic mean of their respective single losses. These global losses are monitored for both the training and validation sets and the hyperparameters of the LSNN are accordingly refined to ensure that these losses can decrease to small values. As a result, the precision of prediction may reach an acceptable level. In the next section, we will evaluate the prediction errors for Partitions $\mathbb 1$, $\mathbb 2$ and $\mathbb 3$.

\section{Validity of LSNN}

The LSNN has been designed based on the GFNN developed in \citet{CT21}, with the specific purpose of applying it to the dynamical model of the PCR3BP. Naturally, we should first evaluate the performance of the LSNN in learning the motion of this system.

In the framework of the PCR3BP, the primary $m_1$ and the secondary $m_2$ are moving on circular orbits about their common centre of mass $O$. The third body, which is massless, is influenced by the gravitational forces of those two massive bodies but does not affect their motion. The motion of all three bodies takes place in the same plane. It is customary to choose non-dimensional parameters such that the primary and the secondary have a total mass of $(m_1 + m_2)=1$, a mutual distance of $a_{12}=1$, and an angular velocity of $n_{12}=1$.

We consider the synodic coordinate system $(x, y)$, which has its origin at $O$ and rotates at a uniform rate $n_{12}$ in the counterclockwise direction. In this coordinate system, the primary and secondary always lie along the $x$-axis at $(-\mu, 0)$ and $(1-\mu, 0)$ respectively. The mass ratio $\mu=m_2/(m_1 + m_2)$ is the only free parameter in the PCR3BP (e.g. the Earth-Moon system, the Sun and a planet, or a binary star). The equations of motion of the particle within any such system can be written as:
\begin{eqnarray}
\ddot{x}-2\dot{y}&=&\frac{\partial U}{\partial x},\nonumber\\
\ddot{y}+2\dot{x}&=&\frac{\partial U}{\partial y},
\label{3b}
\end{eqnarray}
where the ``pseudo-potential'' $U=U(x, y)$ is
\begin{equation}
U=\frac{1}{2}(x^2+y^2)+\frac{1-\mu}{r_1}+\frac{\mu}{r_2}
\label{Ufunction}
\end{equation}and $r_1$ and $r_2$ are the particle's distances to the primary and secondary respectively, given by:
\begin{eqnarray}
r_1^2&=&(x+\mu)^2+y^2,\nonumber\\
r_2^2&=&(x+\mu-1)^2+y^2.
\label{distance}
\end{eqnarray}

It is well known that there is only one integral of the differential system (\ref{3b}), i.e. the Jacobi integral
\begin{equation}
C(x, y, \dot{x}, \dot{y})=2U(x, y)-(\dot{x}^2+\dot{y}^2).
\label{jacobi}
\end{equation} 
In the PCR3BP, the Jacobi integral is equivalent to the Hamiltonian
\begin{equation}
H(q_1,q_2,p_1,p_2)=\frac{1}{2}(p_1^2+p_2^2)+p_1 q_2-p_2 q_1-U^{*},
\label{Hamiltonian}
\end{equation}
where $\bm q=(q_1,q_2)=(x,y)$ is the coordinate, $\bm p=(p_1,p_2)=(\dot{x}-y, \dot{y}+x)$ is the momentum conjugate to $\bm q$, and the $U^{*}$ is the same as $U$ but parameterised by $(q_1,q_2)$. Using equation (\ref{Hamiltonian}), the equations of motion of the massless particle can be written in Hamiltonian form:
\begin{equation}
\dot {\bm q}=+\frac{\partial H}{\partial \bm p},~~~\dot {\bm p}=-\frac{\partial H}{\partial \bm q}.
\label{Hequation}
\end{equation}

In order to generate ground truth orbits for machine learning, we use the Runge-Kutta (RK) method to integrate the equations of motion\footnote{Equations (\ref{3b}) and (\ref{Hequation}) yield identical solutions in terms of $\bm q$ and $\bm p$, despite having different forms.} and record the canonical variables $\bm q$ and $\bm p$. It should be noted that although the equations of motion (\ref{Hequation}) are symplectic, the output orbits of the RK integrator are not strictly symplectic. However, if the RK method is accurate enough to yield a solution that closely approximates the real solution of the Hamiltonian equations of motion, then the RK-generated ground truth orbits can be considered nearly symplectic and possible to be learned well by the LSNN based on the generating function approach. It is worth mentioning that, besides \citet{CT21}, \citet{Grey2019} also used the RK integrator to generate ground truth data for the PCR3BP, which is a non-integrable and non-separable Hamiltonian system. In this work, the focus is on whether the LSNN can provide accurate predictions compared to the ground truth from numerical integration, rather than on the strict symplecticity of the ground truth orbits. The LSNN's performance will also be evaluated for fitting the ground truth orbits generated by the symplectic integrator, as discussed at the end of this paper.

\begin{table*}
\centering
\caption{For the prediction of a test orbit in the PCR3BP, the \textit{partitioned} single losses $loss(\bm Q^j)$ (see equation (\ref{loss1})) for the GFNN proposed by \citet{CT21} and the LSNN developed in this work. The losses are calculated for various partitions, as depicted in Fig. \ref{partition}. Notably, for an extended evolution time of $\ge t_{100}$, we additionally list the losses for Partitions $j\ge\mathbb4$ in this table.}      
\label{3body}
\begin{tabular}{c c c c c}        
\hline                 
 Partition $j$     &  Time interval         &   Network        & $loss(\bm Q^j)$   \\
\hline

\hline

$j =\mathbb1$              &    $[t_{25}, t_{49}]$  &     GFNN         &  0.00248        \\
                   &                        &     LSNN         &  0.00298       \\
\hline
$j =\mathbb2$              &   $[t_{50}, t_{74}]$   &     GFNN         &  0.00805       \\
                   &                        &     LSNN         &  0.00311        \\
\hline
$j =\mathbb3$              & $[t_{75}, t_{99}]$     &     GFNN         &  0.01555        \\
                   &                        &     LSNN         &  0.00272      \\
\hline
\hline
$j =\mathbb4$              &    $[t_{100}, t_{124}]$  &     GFNN         &  0.02349        \\
                   &                        &     LSNN           &  0.00731       \\
\hline
$j =\mathbb5$              &   $[t_{125}, t_{149}]$   &     GFNN         &  0.03227       \\
                   &                        &     LSNN           &  0.00519        \\
\hline
$j =\mathbb6$              & $[t_{150}, t_{174}]$     &     GFNN         &  0.04082
        \\
                   &                        &     LSNN           &  0.00641      \\
\hline
$j =\mathbb7$              &    $[t_{175}, t_{199}]$  &     GFNN         &  0.04936
        \\
                   &                        &     LSNN           &  0.01010       \\

\hline
\end{tabular}
\end{table*}

Now, we briefly review the principal settings in \citet{CT21} for complement.  These are: (1) the primary and secondary have equal masses, i.e. $\mu=0.5$ and they form an inner binary. (2) For the particle, the coordinate is ${\bm q}$ and $\bm p$ is the corresponding conjugate momentum. (3) The particle's orbit starts with $\bm q^{\ast}=(5,0)$ and $\bm p^{\ast} = (0,\frac{1}{\sqrt{4.5}})$ and a time series is recorded during the numerical integration. Then, a subset of these time series points are used as the initial conditions $\bm q_0$ and $\bm p_0$ to build the orbit database for machine learning. This way, $10^5$ and $10^2$ orbits are generated for the training and validation sets respectively. These ground truth orbits are numerically calculated by the 4th-order RK integrator with a step size of $10^{-3}$. (4) For each orbit in the training and validation sets, a time step of $h=0.1$ between two successive time points is adopted.

We notice that the two massive bodies are located at $(-0.5, 0)$ and $(0.5, 0)$, with distances of 0.5 from the origin $O$, while the particle is about ten times farther away from $O$. Due to this hierarchical structure, the triple system can be approximated as a 2-body system, where all the mass is concentrated at the centre of mass $O$ (i.e. the origin) and the particle revolves around $O$. This implies that the evolution of the particle is in the non-chaotic regime of the PCR3BP and its regular motion could be easily learned by the neural network. Another way to see this is through the Jacobi integral. For the adopted initial conditions, the Jacobi integral of the particle is $C^{\ast}=4.896$ (determined by equation (\ref{jacobi})), while the Jacobi integral of the Lagrangian point $L_1$ is $C_{L_1}=4.059$. Since $C^{\ast} > C_{L_1}$, the particle will never visit the primary or secondary, but instead remain on a distant and stable orbit.

Our machine learning experiments are implemented in PYTORCH on CPU/GPU of Google Colaboratory Pro+. Similar to \citet{CT21}, we train the LSNN to learn the modified generating function $S$. The  architecture of the LSNN includes 6 fully connected layers, with 200, 100, 50 and 20 neurons in each hidden layer, respectively. We use the hyperbolic tangent activation function $\tanh$. The batch size is set to be 200 and the Adam optimizer is applied.

As described in Section 2.3, we divide the sequence $(\bm q_i,\bm p_i)~(i\in[0, 99])$ into 4 partitions. We use the initial conditions in the time interval $t\in[t_{0}, t_{24}]$ (i.e. Partition $\mathbb 0$) to predict the particle's motion in the later evolution time of $t\in[t_{25}, t_{99}]$ (i.e. Partitions $\mathbb 1$, $\mathbb 2$, and $\mathbb 3$), and compare our results with those of \citet{CT21}. We start by making predictions for $\text{Partition } \mathbb 1$. When the modified generating function $S^1$ has been learned, we can apply Newton's method to solve the equation:
\begin{equation}
\bm P+\partial_{\bm q} S^1(\bm q,\bm P)-\bm p=0,
\label{newton}
\end{equation}
where we set $\bm p=\bm p_0$, $\bm q=\bm q_0$. Then, we can solve for the predicted coordinate $\tilde{\bm p}_{25}=\bm P$. Using the coordinate $\bm P$, we can calculate the value of $\bm Q$ from equation
\begin{equation}
\bm Q=\bm q+\partial_{\bm P} S^1(\bm{q},\bm{P}).
\end{equation}
Thus, we can obtain the predicted momentum $\tilde{\bm q}_{25}=\bm Q$. In the next step we set $\bm p=\bm p_1$ and $\bm q=\bm q_1$ and use the learned modified generating function $S^1$ to obtain $\bm P=\tilde{\bm p}_{26}$ and $\bm Q=\tilde{\bm q}_{26}$. We repeat this process by updating $\bm p=\bm p_i$ and $\bm q=\bm q_i$ at each step for $i=2,3,\ldots,24$, until we obtain the final predicted variables  $\tilde{\bm p}_{49}$ and $\tilde{\bm q}_{49}$. This completes the symplectic mappings (\ref{P1}) for $\text{Partition } \mathbb 1$.

Independently, we then make predictions on $\text{Partition } \mathbb 2$. We again use $\text{Partition } \mathbb 0$ as the initial data and follow the same procedure as above to generate the symplectic mappings (\ref{P2}). The symplectic mappings (\ref{P3}) for $\text{Partition} \mathbb 3$ can be obtained as well in a similar fashion.

After training our LSNN, we evaluate its performance on predicting the same test orbit used in \citet{CT21}, which starts with the initial condition of $\bm q_0=(5.01,-0.01)$ and $\bm p_0= (0,\frac{1}{\sqrt{4.51}})$. For this test orbit, we calculate the \textit{partitioned} single losses $loss(Q^j)$ (see equation (\ref{loss1})) of both our best-trained LSNN and the GFNN\footnote{We adopt the codes used in \citet{CT21} on the website: \url{https://proceedings.mlr.press/v139/chen21r.html}, and then apply equation (\ref{loss1}) to get the losses of the GFNN presented in Table \ref{3body}.}. As shown in Table \ref{3body}, our LSNN achieves very small losses across all partitions $\mathbb{1}$-$\mathbb{3}$, at the level of $\sim0.003$. Comparatively, the GFNN exhibits a similar loss of around 0.003 during the early evolution time of Partition $\mathbb{1}$ (i.e. $t\in[t_{25}, t_{49}]$). However, the GFNN's losses increase to larger values of about 0.008 and 0.015 for Partition $\mathbb{2}$ (i.e. $t\in[t_{50}, t_{74}]$) and Partition $\mathbb{3}$ (i.e. $t\in[t_{75}, t_{99}]$), respectively. From this point of view, our LSNN's predictions demonstrate comparable or even smaller losses compared to those of the GFNN.

\begin{figure}
 \hspace{0cm}
  \centering
  \includegraphics[width=9cm]{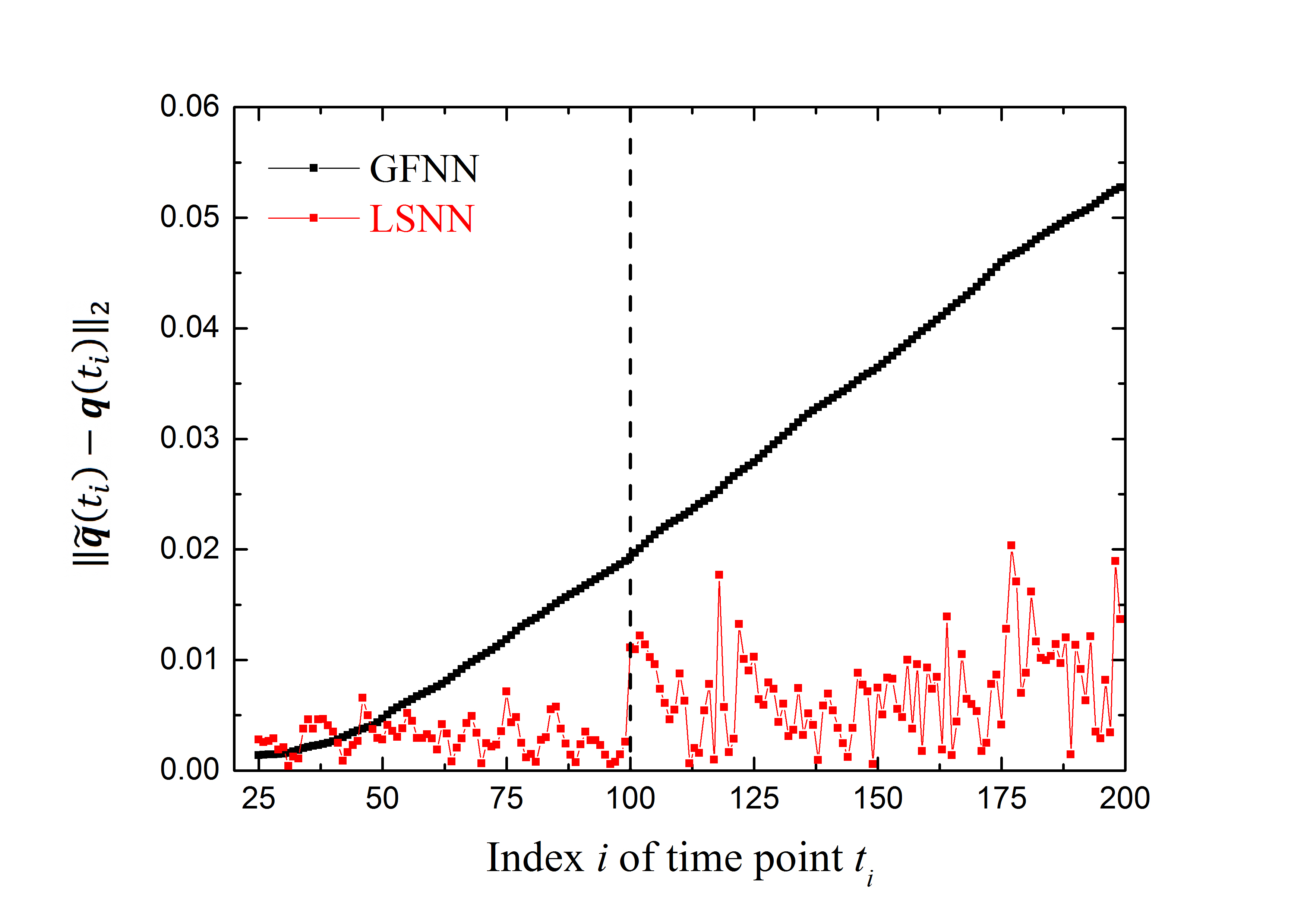}
\caption{For the same prediction of the test orbit as in Table \ref{3body}, the difference $||\tilde{\bm q(t_i)} - \bm q(t_i)||_2$ (see equation (\ref{2-norm})) at individual time points $t_i$ ($i=25$-199). Here, $\tilde{\bm q}$ is the predicted coordinate and $\bm q$ is the ground truth. The red curve refers to our LSNN, while the black one refers to the GFNN of \citet{CT21}. For reference, a vertical (dashed) line is plotted at the time point of $t_{100}$. To the left of this line, Partitions $\mathbb 1$, $\mathbb 2$ and $\mathbb 3$ have relatively smaller errors, and they would be used to predict the evolution of the KBOs in Section 4.}
  \label{GFNN}
\end{figure}

In order to visualize the performance of our LSNN better, we plot in Fig. \ref{GFNN} the difference between the prediction of the coordinate $\tilde{\bm q}(t_i)$ and the ground truth $\bm q(t_i)$, measured by the Euclidian 2-norm:
\begin{equation}\label{2-norm}
||\tilde{\bm q} - \bm q||_2=\sqrt{(\tilde q_x- q_x)^2+(\tilde q_y- q_y)^2}. 
\end{equation}
It is evident that for the time periods from $t_{25}$ to $t_{99}$ (to the left of the vertical line), the prediction errors of our LSNN (the red curve) are consistently small. However, the prediction errors of the GFNN (the black curve) tend to increase more prominently over time and become relatively larger at the time point $t_{99}$. Therefore, it seems that our LSNN has a rate of error growth that is smaller than that of the GFNN for the considered evolution time (i.e. for Partitions $\mathbb 1$-$\mathbb 3$).

Nevertheless, it would be tempting to see the performance of our LSNN if the evolution time $t_i$ of the system is extended. Hence, we consider another four following partitions starting from $t_{100}$, denoted by Partitions $\mathbb 4$-$\mathbb 7$, and each of these partitions contains 25 time points (see Table \ref{3body}). This allows us to make a further comparison against \citet{CT21} up to the time of $t_{199}$. As shown in Fig. \ref{GFNN}, in the extended time interval of $[t_{100}, t_{199}]$ (i.e. to the right of the vertical line), although the mean error of our LSNN appears to increase apparently over time, the losses $loss(\bm Q^j)$ of our LSNN are still relatively smaller than those of the GFNN.

We want to stress that extending the learning of the particle's evolution from $t_{100}$ to $t_{199}$ by using our partition approach is a non-trivial task. As we demonstrated earlier, for either Partition $\mathbb 2$ or $\mathbb 3$, the learning process is a simple repetition of what we did for Partition $\mathbb{1}$. However, due to the intense variation of the training data in the time interval of $t\ge t_{100}$, we have to introduce some delicate translations for Partitions $\mathbb{4}$-$\mathbb{7}$ in order to develop the LSNN. To accomplish this, we write a revised generating function
\begin{equation}\label{gf2}
F(q,P) = (q+d_1)(P-d_2) + S(q,P), 
\end{equation}
where $(Q, P)$ stands for a point in Partition $\mathbb 4$-$\mathbb 7$, $(q, p)$ is point in Partition $\mathbb 0$, and the parameters $d_1$ and $d_2$ measure the magnitude of the translation. Let $P^{\prime}=P-d_2$ and $Q^{\prime}=Q-d_1$. From the generating function (\ref{gf2}), we can obtain 
\begin{eqnarray}
p = P^{\prime}+\frac{\partial{S}}{\partial q},\nonumber\\
Q^{\prime} = q +\frac{\partial{S}}{\partial P},
\label{gf2pq}
\end{eqnarray}
which represents the symplectic mapping 
\begin{equation}\label{sp2}
(q,p)\rightarrow(Q^{\prime},P^{\prime}).
\end{equation}
By adopting appropriate values of $d_1$ and $d_2$, the translated point $( Q^{\prime}, P^{\prime})$ could be close to the point $(q, p)$ in Partition $\mathbb 0$. This enabled our LSNN to learn the symplectic mapping (\ref{sp2}) with high precision.

Theoretically speaking, we could learn the orbit evolution further in time, e.g. $t\ge t_{200}$, but the amplitude of the variation of the training data would become even larger and then the above translation (i.e. the values of $d_1$ and $d_2$) would be extremely difficult to implement. In practice, the considered evolution time up to $t_{199}$ is long enough for the purposes of this study. Since the prediction of the test orbit has relatively smaller errors at the time of $t_{25}$-$t_{99}$, only the first three partitions $\mathbb 1$, $\mathbb 2$ and $\mathbb 3$ will be used to predict the evolution of the KBOs in the next section.

\section{Applications}

Beyond the orbit of Neptune, there exists a disk of small bodies orbiting the Sun known as the Kuiper belt. By now, thousands of the KBOs have been observed and they can be classified according to their orbital characteristics: the resonant, the classical (i.e. non-resonant), the scattered, and the detached objects. Studying the dynamical evolution of the KBOs may provide important clues about the early history of our Solar System. 

We will now utilise the designed LSNN to learn the behaviours of the KBOs outside or in the mean motion resonance (MMR) with Neptune. In the framework of the PCR3BP, we refer to the Sun with mass $m_1$ as the primary, the planet Neptune with mass $m_2$ as the secondary and a KBO as the massless third body. Therefore, in the equations of motion of the KBO particle, i.e. equation (\ref{3b}), the mass ratio is $\mu\approx5.146\times 10^{-5}$. Next, we generate the training and validation data by numerical simulations. As discussed at the end of Section 3, for the orbit of each KBO, we take 100 equally spaced time points for the data in Partition $\mathbb 0$ (for the initial condition) and in Partitions $\mathbb 1$-$\mathbb 3$ (for the prediction).

\subsection{Non-resonant KBOs}

\begin{figure}
 \hspace{0cm}
  \centering
  \includegraphics[width=9cm]{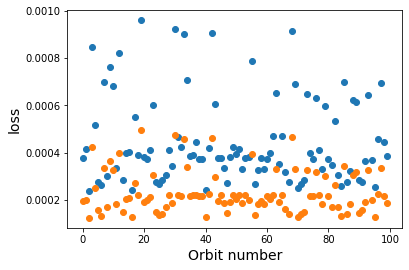}
\caption{Learning of the motion of non-resonant KBOs by the LSNN: the single losses of the predicted orbits of the 100 validation samples for the time step of $h=0.1$. The blue dots refer to the coordinate losses $loss(\bm Q)$ and the yellow dots indicate the momentum losses $loss(\bm P)$, as defined in equation (\ref{loss3}).}
  \label{loss_KBOnon}
\end{figure}

The classical KBOs have semimajor axes roughly between 42 AU and 48 AU and are defined as objects outside Neptune's MMRs. In the PCR3BP, we consider the particles with initial non-dimensional semimajor axes $a=1.44$. Since the distance of $\sim 30$ AU between the Sun and Neptune is set to be 1, the particles' actual semimajor axes are about 43.2 AU. At this location, there are no MMRs of 7th- or lower orders \citep{Li2023}. To ensure that the KBO particles are non-resonant, we place them on nearly circular orbits with eccentricities $e=0.01$. As for the orbital angles, the initial longitudes of perihelia $\varpi$ are set to be $60^{\circ}$, and the initial mean longitudes $\lambda$ are randomly selected between 0 and $360^{\circ}$. 

As in Section 3.1, the ground truth orbits are generated by using the 4th-order RK integrator. The training and validation sets consists of $10^5$ and $10^2$ orbits, respectively. For each orbit, there are 100 time points with a time step of $h=0.1$, so the total evolution time is 10 in non-dimensional units. According to Kepler's third law, the KBO particle has an orbital period of $\sqrt[3]{1.44/1}\times2\pi\approx7.1$, indicating that it could complete more than one full orbit during the evolution time of 10. We note that in the PCR3BP comprising the Sun, Neptune and a KBO, the dimensionless time unit is $2\pi$, which corresponds to the orbital period of Neptune moving around the Sun, i.e. 164.79 yr. Therefore, the actual evolution time is equal to $(10/2\pi)\times164.79$ yr $\approx262.45$ yr. A similar transformation is also used to calculate the actual evolution time of the resonant KBOs in Section 4.2.  

We then proceed to re-train our LSNN model using the partitioned data and the results are presented in Fig. \ref{loss_KBOnon}. For each of the 100 trajectories from the validation set, we plot the single losses of the predicted coordinates $\tilde{\bm q}$ (blue dots) and momenta $\tilde{\bm p}$ (yellow dots),  i.e. $loss(\bm Q)$ and $loss(\bm P)$ defined in equation (\ref{loss3}). We notice that the maximal loss, which is associated to $loss(\bm Q)$, is less than 0.0009, while the losses $loss(\bm P)$ are generally smaller, ranging between 0.0001 and 0.0005. As we demonstrated in Section 2.3, in order to make the symplectic transformations via the modified generating function $S(\bm{q},\bm{P})$, we first derive the new momentum $P$, then the new coordinate $\bm Q$ is predicted by $\bm Q=\bm q+\partial_{\bm P} S(\bm{q},\bm{P})$. Therefore, the loss $loss(\bm Q)$ should naturally encompass the loss $loss(\bm P)$ and this explains why $loss(\bm Q)$ is larger than $loss(\bm P)$ in Fig. \ref{loss_KBOnon}. Despite this, the LSNN appears to work better in predicting the evolution of the non-resonant KBOs than in the case of learning the "binary + particle" system in Section 3, where $loss(\bm Q)$ is an order of magnitude larger.

\subsection{Resonant KBOs} 

The population of resonant KBOs provides strong evidence for the orbital migration of the Jovian planets. The study of the resonant KBOs is very important for us in understanding the early shake-up of the outer Solar System, and a lot of research has been carried out on this topic since the early 1990s \citep[e.g.][]{malh93, malh95, malh05, Li2014, nesy16, pike17, lawl19}. In our recent work \citep{li22}, we introduced a machine learning method that can quickly predict the orbital evolution of the KBOs in the 2:3 MMR with Neptune. In the framework of the PCR3BR, due to the restrictions of the invariant tori, the motions of the 2:3 resonant KBOs are regular and can be learned quite accurately.

As mentioned in Section 1, we would like to use the coordinate $\bm q$ and momentum $\bm p$ as training data because such variables can be applied to the evolution of a wide range of physical systems. However, by learning the orbits in terms of $\bm q$ and $\bm p$, our previous ANN designed in \citet{li22} can neither keep the the resonator's evolution symplectic nor conserve the Jacobi integral well. In what follows, we will investigate the performance of the newly developed LSNN in predicting the time series $(\bm q_i, \bm p_i)$ for the 2:3 resonant KBOs.

\begin{figure}
 \hspace{0cm}
  \centering
  \includegraphics[width=8.5cm]{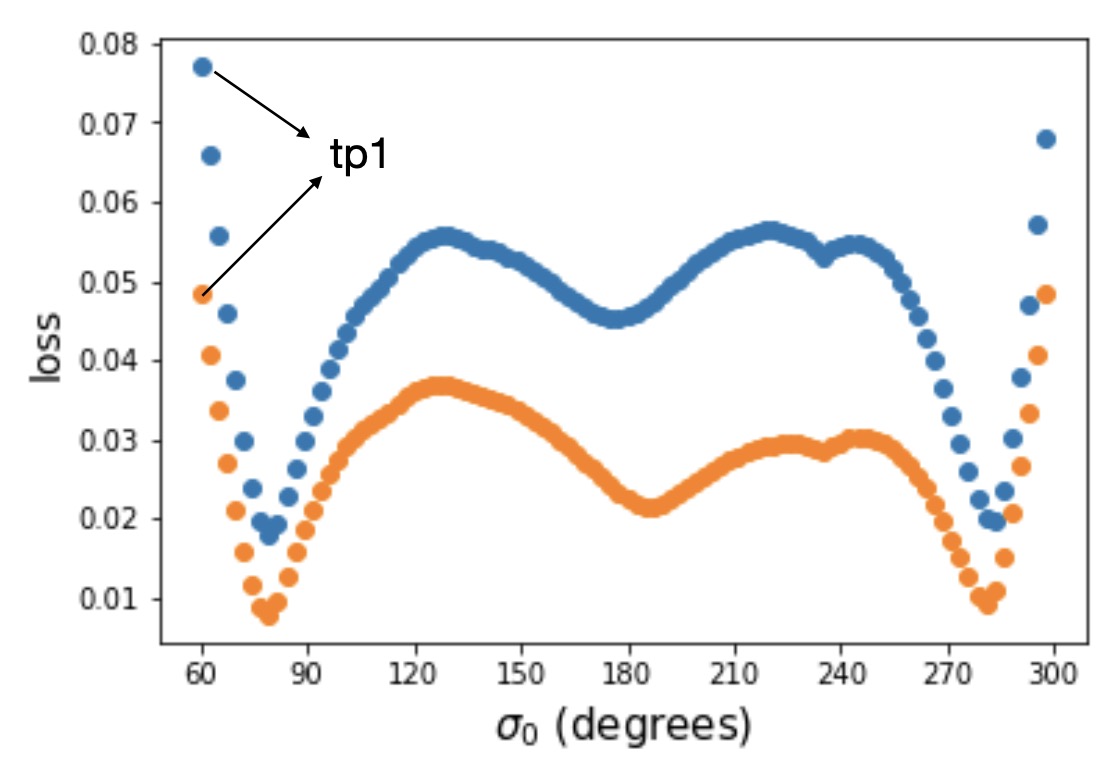}
\caption{Similar to Fig. \ref{loss_KBOnon}, but for the prediction of the motion of resonant KBOs by the LSNN. Here, we adopt a very large time step of $h=9.65$ to cover a total evolution time of 25000 yr, which is longer than a full libration period of the 2:3 resonant KBOs. The horizontal axis shows the initial resonant angle $\sigma_0$, which ranges from $60^{\circ}$ to $300^{\circ}$, centred at the stable libration centre of $180^{\circ}$.}
  \label{e01_profile}
\end{figure}

For the resonant particles, their initial conditions are exactly the same as we used in \citet{li22}:  $a_0=1.31$ (the nominal location of the 2:3 resonance), $e_0=0.1$ (to ensure regular motion throughout phase space), $\varpi_0=60^{\circ}$, and the initial resonant angles $\sigma_0$ in the range of $[180^{\circ}-120^{\circ}, 180^{\circ}+120^{\circ}]$, where $180^{\circ}$ is the stable libration centre. Then, the initial value of $\lambda$ can be determined from a given resonant argument $\sigma(=\sigma_0)$ by the equation:
\begin{equation}
\sigma=3\lambda-2\lambda_N-\varpi,
\label{ResAng}
\end{equation}
where $\lambda_N$ is the mean longitude of Neptune. We note that $\varpi_0$ can have an arbitrary value since $\sigma$ is invariant to changes in $\varpi$. For a given set of the orbital elements $(a, e, \varpi, \lambda)$, the initial conditions for the 2:3 resonators are transformed in terms of the coordinate $\bm q$ and momentum $\bm p$. Then, we numerically propagate the equations of motion to generate the training and validation orbits. To be consistent with \citet{li22}, we also employ here the 8th-order RK integrator with a time step of $\sim0.01$ yr and a local error tolerance of $10^{-20}$. Additionally, for the numerical integration over the long timescale that we are about to consider below, the ground truth obtained from the 8th-order RK integrator can be more accurate than that from the 4th-order one and closer to the real symplectic orbits.

To  effectively capture the behaviour of the 2:3 resonant KBOs, the time evolution needs to span at least a full libration period, which is approximately 20000 years, corresponding to a non-dimensional time of $\sim762.5$. This is a significant increase in timescale compared to the two 3-body models considered previously, where the evolution time was only 10. In order to fulfill this task, we have increased the time step $h$ to 9.65. Then the 100 time points lead to a total evolution time of 965, which is equivalent to the real timespan of 25000 yr that we adopted in \citet{li22}. It should be noted that the LSNN uses Partition $\mathbb 0$ as the initial condition to predict Partitions $\mathbb 1$, $\mathbb 2$ and $\mathbb 3$. This approach is also consistent with our previous work, where we utilised the known first 1/4 period of the orbit to predict the subsequent 3/4 period data.

\begin{figure*}
  \centering
  \begin{minipage}[c]{1\textwidth}
  \vspace{0 cm}
  \includegraphics[width=9cm]{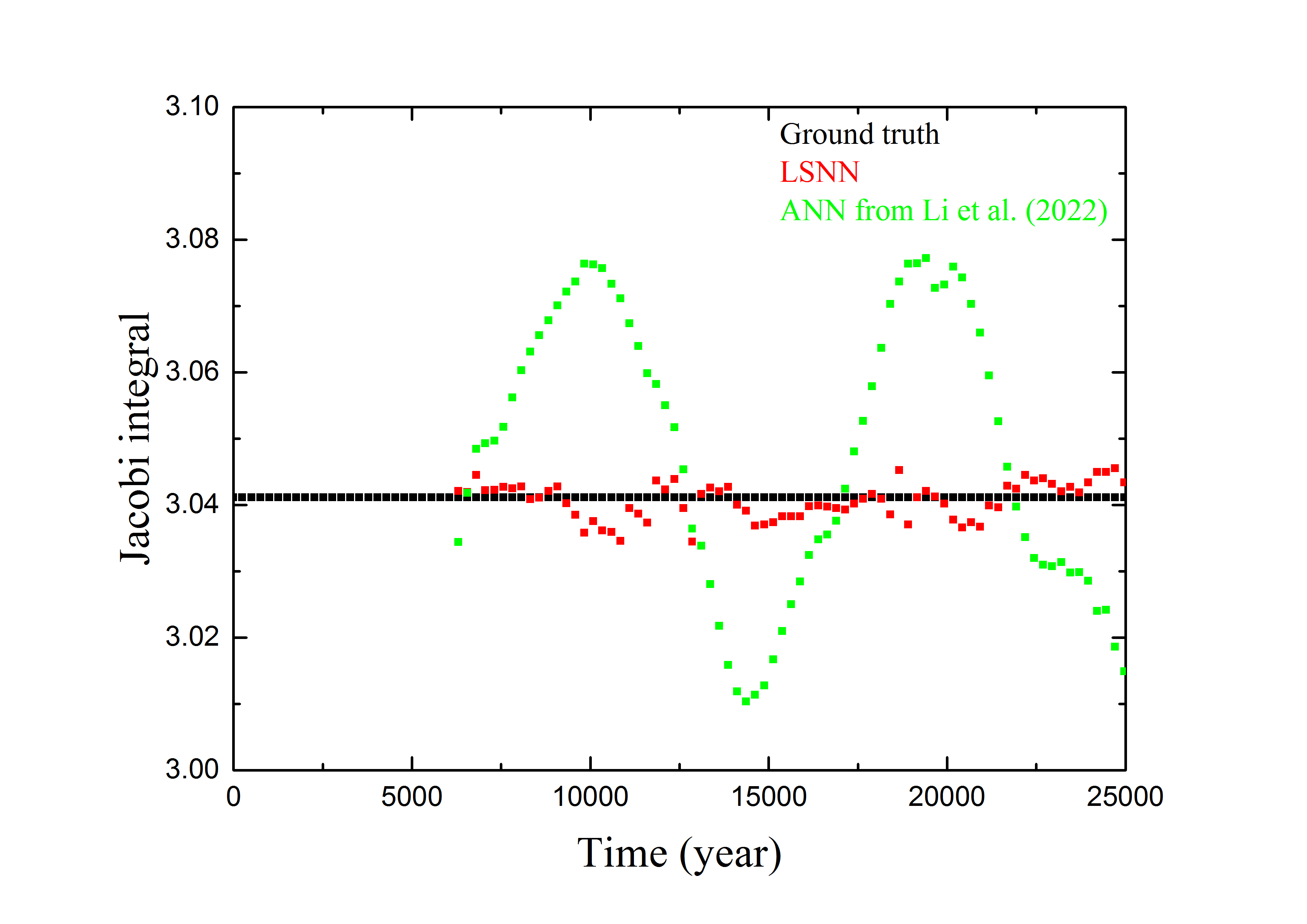}
  \end{minipage}
  \begin{minipage}[c]{1\textwidth}
  \vspace{0 cm}
  \includegraphics[width=9cm]{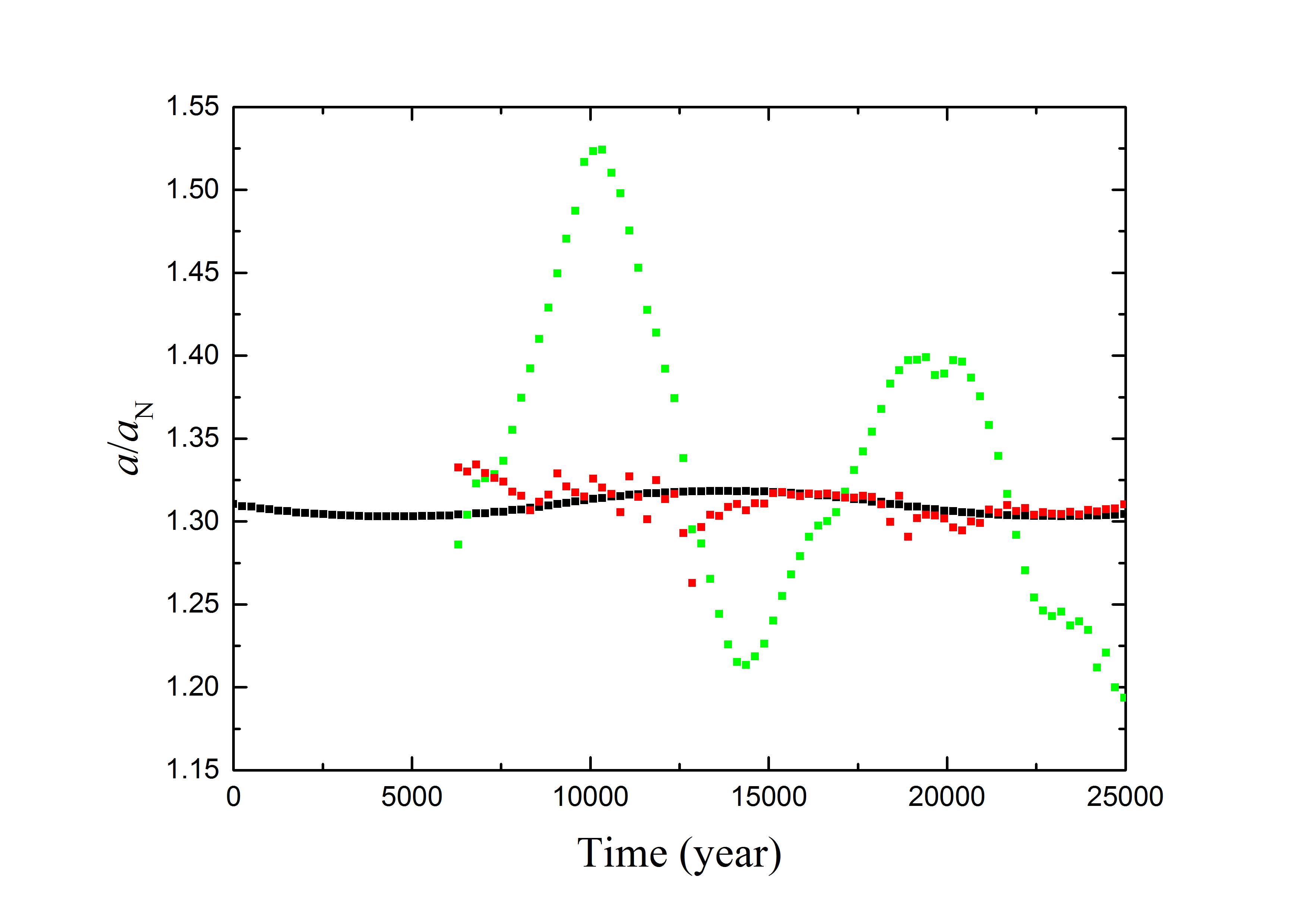}
  \includegraphics[width=9cm]{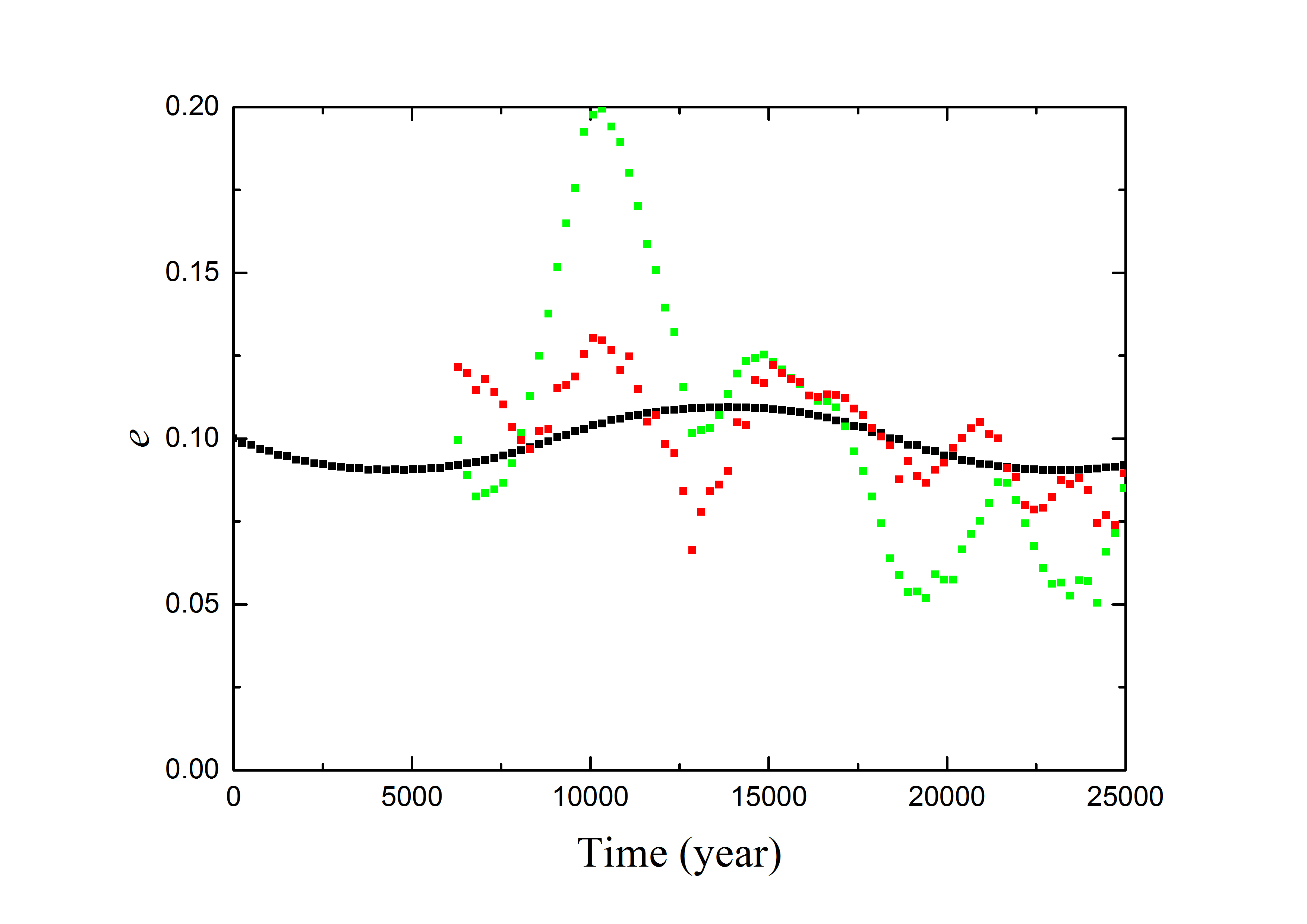}
  \end{minipage}
  \begin{minipage}[c]{1\textwidth}
  \vspace{0 cm}
  \includegraphics[width=9cm]{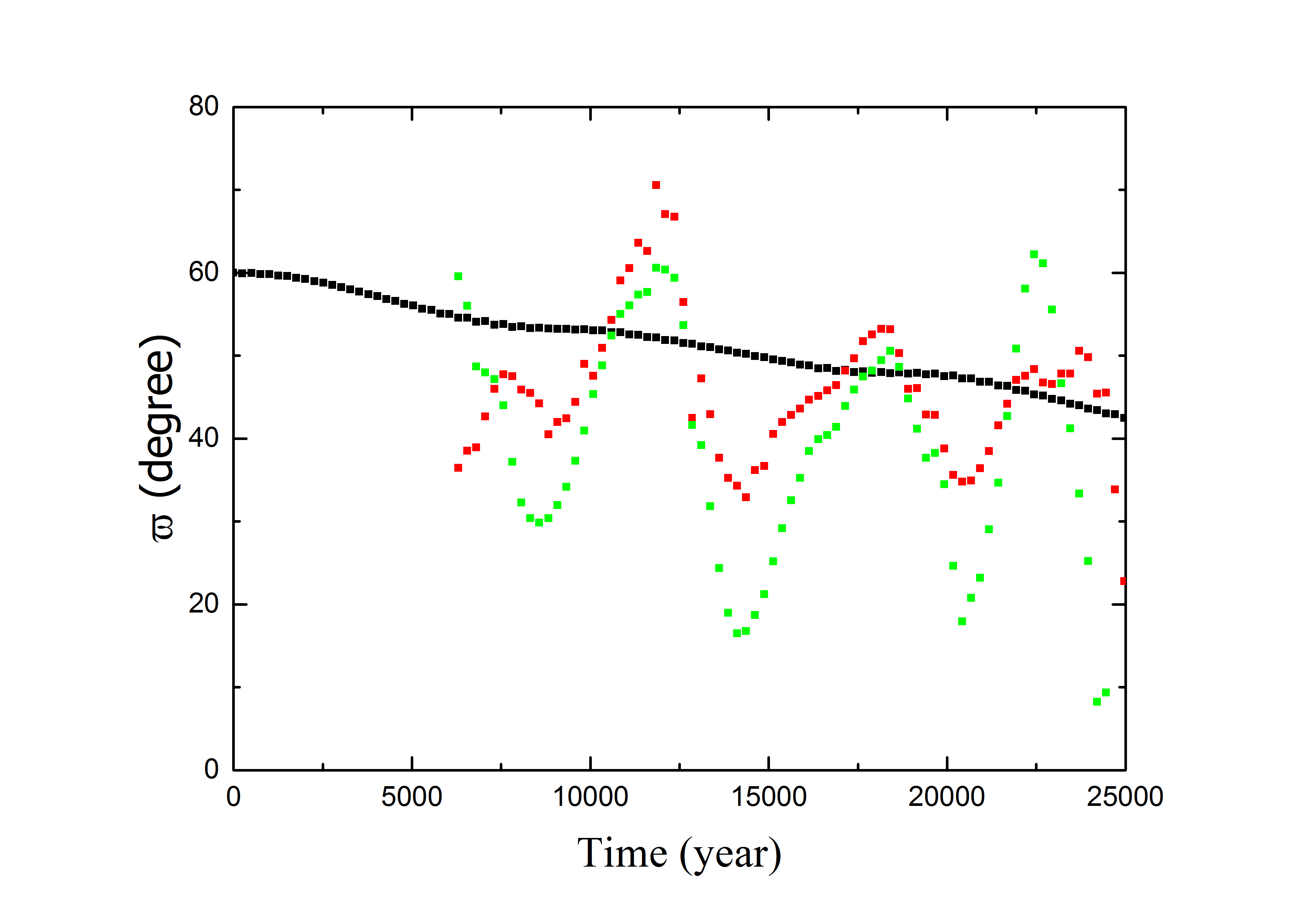}
  \includegraphics[width=9cm]{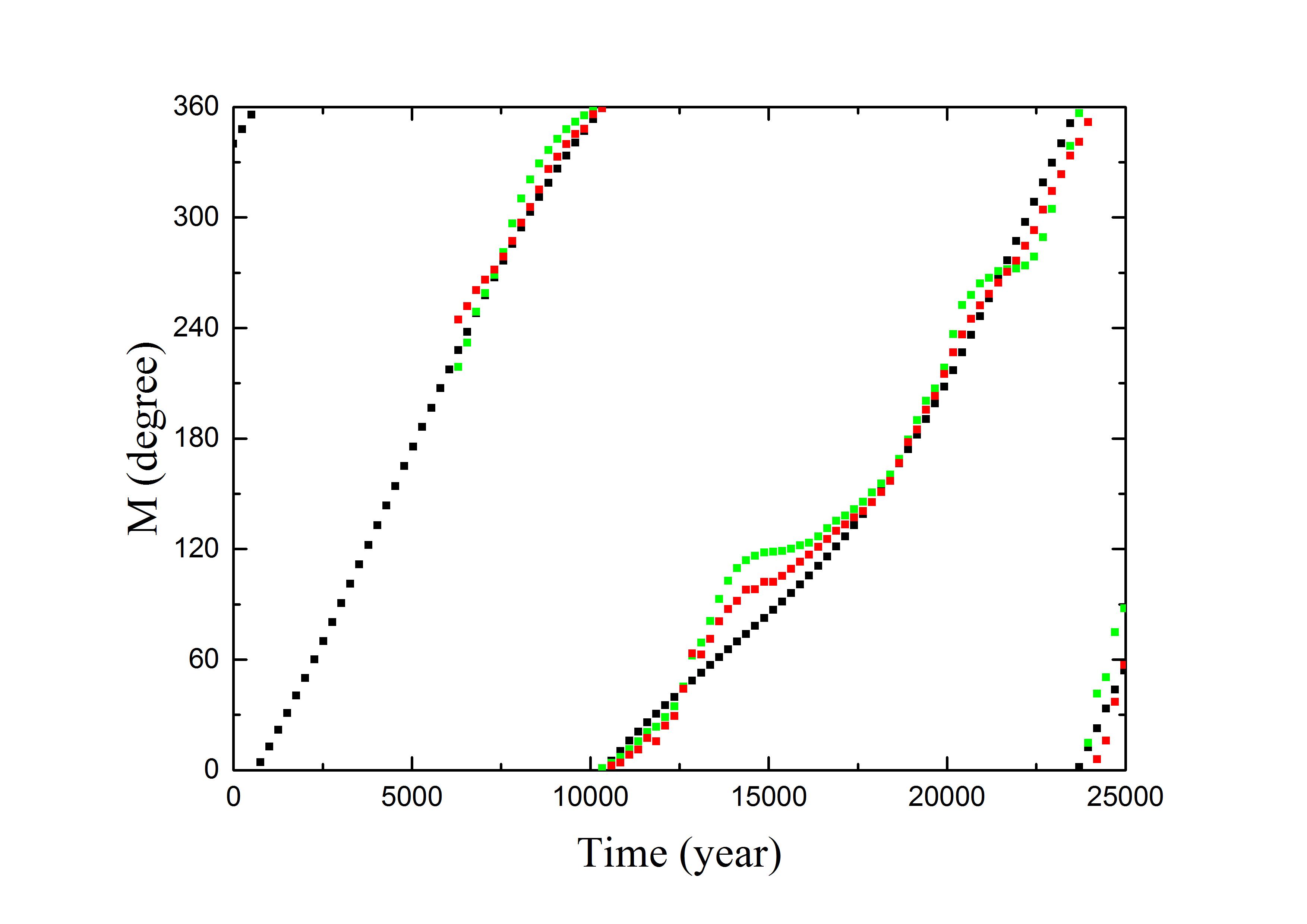}
  \end{minipage}
   \vspace{0 cm}
  \caption{For the validation sample tp1 starting with $\sigma_0=60^{\circ}$ (see Fig. \ref{e01_profile}), the temporal evolution of the Jacobi integral (upper panel) and the orbital elements $a$, $e$, $\varpi$, $M$ (lower four panels). The black dots refer to the ground truth calculated by the numerical simulations. Provided the first 1/4 period of the ground truth orbit as the initial condition, the following 3/4 period orbits are predicted by the newly developed LSNN (red dots) and the previous ANN constructed in \citet{li22} (green dots). In the framework of the PCR3BP, the particle's semimajor axis, $a$, is given in units of Neptune's semimajor axis $a_N\sim30$ AU.}
 \label{stability}
\end{figure*} 

Similar to the LSNN designed for the non-resonant KBOs, but using a much larger time step $h$ for the time series $(\bm q_i, \bm p_i)$, our newly developed machine learning method is further applied to the 2:3 resonant KBOs. The single losses on the validation set for 100 orbits with different $\sigma_0$ are presented in Fig. \ref{e01_profile}. The blue and yellow dots indicate $loss(\bm Q)$ and $loss(\bm P)$ respectively, which are calculated from equation (\ref{loss3}) for the 75 predicted time series points $(\tilde{\bm q}_{25}, \tilde{\bm p}_{25}),\dots,(\tilde{\bm q}_{99}, \tilde{\bm p}_{99})$. We notice that: (1) the maximum $loss(\bm Q)\sim0.08$ and $loss(\bm P)\sim0.05$ are at both ends of the curves; (2) there are two valleys where the losses reach a minimum of about $\sim0.01$; (3) there are two peaks as local maxima. The profile of the loss curve looks very similar to that in Fig. 3 from \citet{li22}, as well as the magnitude of the loss. However, we cannot directly compare the performance of the LSNN with that of the ANN in our previous work. This is because the data sequences here consist of coordinates and momenta, but not of orbital elements.

To demonstrate the good performance of the LSNN, we select the validation sample ``tp1'' starting with $\sigma_0=60^\circ$, which has the largest losses of $loss(\bm Q)\sim0.08$ and $loss(\bm P)\sim0.05$ (as shown in Fig. \ref{e01_profile}). Fig. \ref{stability} presents the evolution for the sample tp1, with associated orbital parameters that can be simply obtained from its coordinate and momentum variables. The black dots represent the ground truth values calculated by numerical integration and the red dots denote the predictions made by the LSNN for the time period from $t_{25}$ to $t_{99}$, where $t_{99}$ corresponds to the real time of 25000 yr. For comparison, we also apply the ANN constructed in \citet{li22} to make additional predictions and the results are indicated by the green dots.

The upper panel of Fig. \ref{stability} shows the time evolution of the Jacobi integral $C$ for the sample tp1. As can be seen, the predictions made by our previous ANN (green dots) seem to considerably deviate from the ground truth (black dots), suggesting that the conservation of $C$ can hardly hold in this situation. But with regard to the LSNN, the approximate values of $C$ (red dots) differ from the ground truth only by less than 0.007 at any given time point. Therefore, the first improvement of the LSNN on predicting the motion of the 2:3 resonators is that the conservation of the Jacobi integral can be well maintained.

As shown in the four lower panels of Fig. \ref{stability}, the second improvement is the increased accuracy of the predictions for the orbital elements $a$, $e$, $\varpi$ and $M$ ($M=\lambda-\varpi$ is the mean anomaly). Compared to the predictions from our previous ANN (green dots), the LSNN predictions (red dots) are much closer to the ground truth values (black dots), particularly for the slow variables $a$ and $e$. In this sense, the LSNN performs better at learning and predicting the orbit evolution in terms of the canonical variables $\bm q$ and $\bm p$ of the Hamiltonian system.  

We must point out that particle tp1 has the maximum losses; there are other particles, however, from the validation set that achieve considerably smaller losses (e.g. at $\sigma_0\sim90^\circ$, $180^\circ$ and $270^\circ$ as shown in Fig. \ref{e01_profile}) and thus the predictions of their orbits would be much more accurate. In summary, our LSNN can work well even when the time step $h$ of the time series $(\bm q_i, \bm p_i)$ is relatively large, and this is the main advantage that we want to highlight.


\section{Conclusions and discussion}

In our previous study, we investigated the machine learning predictions for a specific Hamiltonian system of the regular resonant motion, where the particle's evolution was learned in terms of the orbital elements \citep{li22}. As a generalisation, we aim at extending this approach by using the training data consisting of the coordinate $\bm p$ and the conjugate momentum $\bm q$, which enables the machine learning technique to be applied to more general Hamiltonian systems. As a matter of fact, \citet{li22} attempted to train a fully connected ANN to learn the time series $(\bm q(t), \bm p(t))$, however, the Jacobi integral was not well conserved for the predicted orbits. To solve this issue, we propose to develop the ANN further to maintain the symplecticity of the Hamiltonian system's evolution.

\begin{figure*}
  \centering
  \begin{minipage}[c]{1\textwidth}
  \vspace{0 cm}
  \includegraphics[width=9cm]{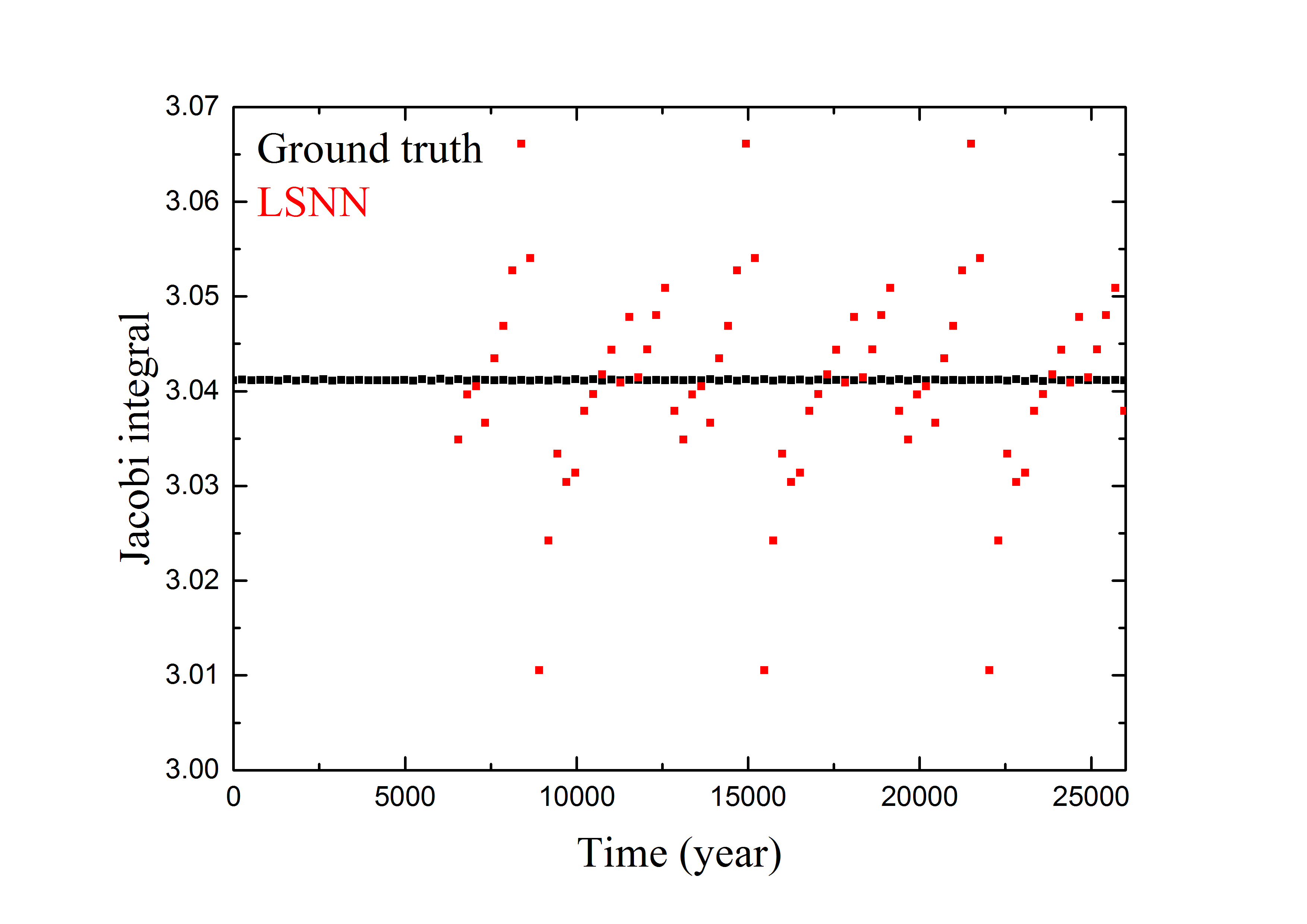}
  \includegraphics[width=9cm]{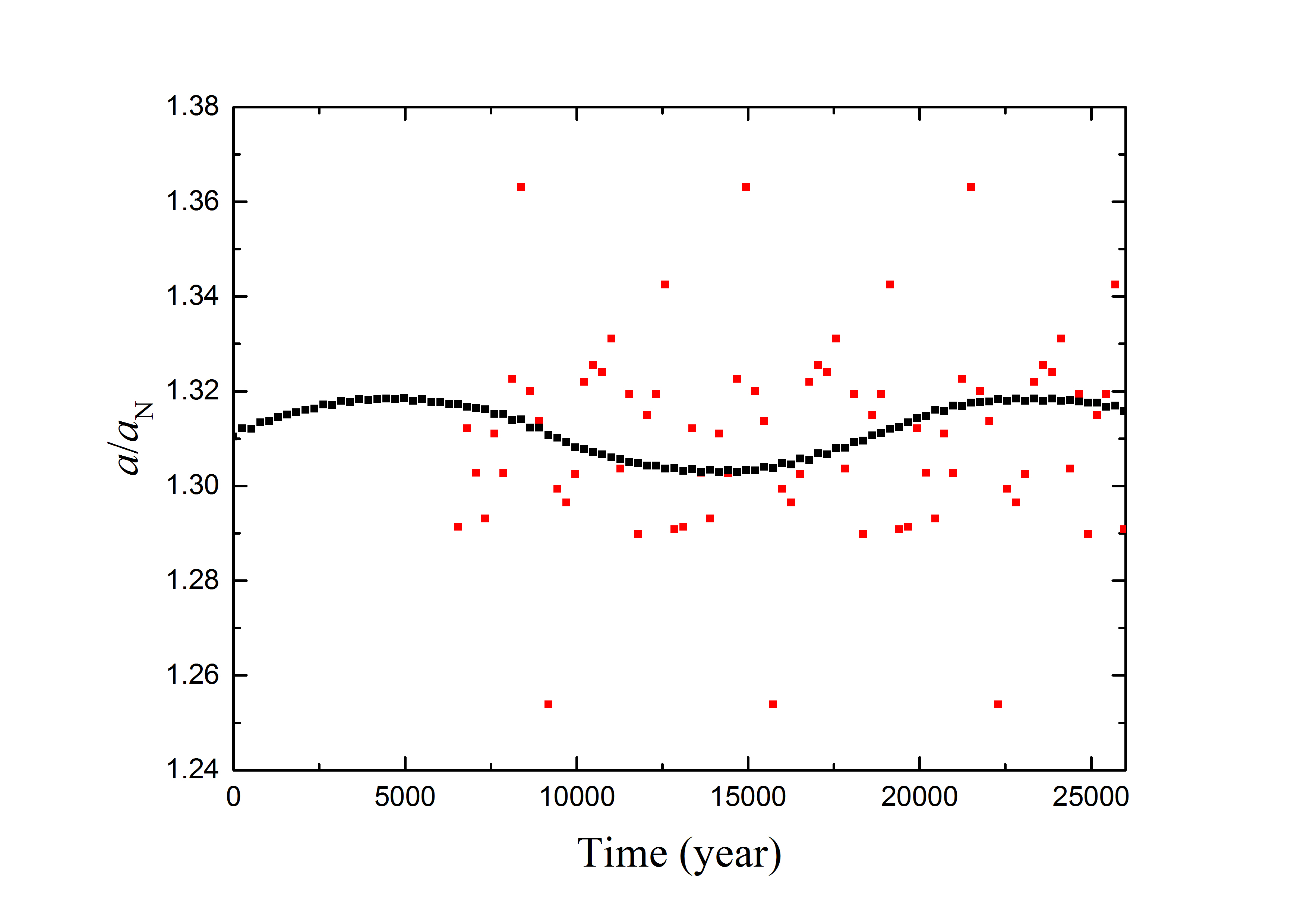}
  \end{minipage}
  \caption{Similar to Fig. \ref{stability}, but for the LSNN trained and validated on data generated from the symplectic integrator instead of the RK method. Only the predictions of the Jacobi integral (left panel) and the particle's semimajor axis $a$ (right panel) are presented for illustrating the failure of our LSNN when using the symplectic integrator to generate the ground truth orbits.}
 \label{SymInt}
\end{figure*} 

In this work, we draw inspiration from the GFNN proposed by \citet{CT21} in order to construct a machine learning model for the symplectic mapping which is represented by a generating function. When we consider a time series comprising discrete points $(\bm q_i, \bm p_i)=(\bm q(t_i), \bm p(t_i))$ ($t_i\in [t_0, t_{99}]$), in mathematics, the generating function can naturally lead to the symplectic transformation from one point to another. As a development in this paper, we divide the 100 time series points into 4 partitions: $\text{Partition } \mathbb 0=\{(\bm q_0, \bm p_{0}), \dots, (\bm q_{24}, \bm p_{24})\}$, $\text{Partition } \mathbb 1=\{(\bm q_{25}, \bm p_{25}) \dots, (\bm q_{49}, \bm p_{49})\}$, $\text{Partition } \mathbb 2=\{(\bm q_{50}, \bm p_{50}) \dots, (\bm q_{74}, \bm p_{74})\}$, and $\text{Partition } \mathbb 3=\{(\bm q_{75}, \bm p_{75}) \dots, (\bm q_{99}, \bm p_{99})\}$. Then we build symplectic mappings from the initial condition of $\text{Partition } \mathbb 0$ to each of the $\text{Partitions } \mathbb 1$, $\mathbb 2$ and $\mathbb 3$ by the modified generating functions $S^1$, $S^2$ and $S^3$ respectively. After $S^1$, $S^2$ and $S^3$ have been learned by the neural network, we obtain three independent groups of symplectic mappings (\ref{P1}), (\ref{P2}) and (\ref{P3}). This new machine learning structure is called large-step neural network (LSNN).

The designed LSNN is first applied to the same PCR3BP model as in \citet{CT21}, where a massless particle orbits around a binary with equal masses. For learning the time series $(\bm q_i, \bm p_i)$ from time $t_{25}$ to $t_{99}$, we adopt the same time step of $h=0.1$ as the one used by the GFNN. The results show that our LSNN has rather small prediction errors over the considered time period, and it has a rate of error growth that is lower compared to that of the GFNN. We then argue that the performance of the LSNN is sufficiently accurate to provide orbit predictions. We also note that this result is further validated for a longer evolution time up to $t_{199}$. However, for an even longer evolution time (e.g. $t=t_{1000}$ considered in \citet{CT21}), we have not yet determined the validity of the LSNN due to some technical issues. Nevertheless, since the LSNN can deal with the case of a large step size $h$ between two successive time points $t_i$ and $t_{i+1}$, quite long-term predictions are allowed.

As practical applications of the LSNN, we have evaluated its predictions on the behaviours of KBOs beyond the orbit of Neptune. More specifically, we investigated the two cases of the non-resonant KBOs and the 2: 3 resonant KBOs:

(1) For the non-resonant KBOs, their orbital periods are about 7.1 in the non-dimensional units. Thus, as in the above case of ``binary + particle", a small time step of $h=0.1$ is adopted to train the LSNN on the 100 time series points $(\bm q_i, \bm p_i)$, which correspond to an evolution time of 10, longer than the orbital period of 7.1. As expected, given such a small $h$, our LSNN can achieve a highly accurate prediction with losses of the order of 0.0001. 

(2) For the 2:3 resonant KBOs, the evolution over a full libration cycle of about 20000 yr needs to be considered. Thus, we increased the time step $h$ from 0.1 to 9.65, leading to a total evolution time of 25000 yr as used in our previous study \citep{li22}. Despite the significantly larger time step (i.e. almost 100 times larger), our LSNN still performs very well. Compared to the previous ANN model constructed in \citet{li22}, the LSNN shows two major improvements: (i) it conserves the Jacobi integral at a much higher level; (ii) it predicts the orbital elements $a$, $e$, $\varpi$, and $M$ with much higher accuracy, even for the validation samples with maximum losses.

Taken as a whole, the main advantage of the newly constructed LSNN is its potential ability to predict long-term evolution and conserve energy for future applications in general Hamiltonian systems by learning the coordinate and momentum variables. It should be noted that our LSNN is still based on the GFNN. We actually have made significant adjustments to the structure of the learning data, but we did not contribute the improvement to the network design itself.

For numerically integrating the Hamiltonian equations of motion (\ref{Hequation}) that describe the symplectic evolution of the particle, the non-symplectic RK integrator was used. We have already stated that, as long as the RK method is accurate enough, its solution could be very close to the symplectic solution of the original Hamiltonian system. This approximation allows our LSNN to successfully learn the generating function from the RK ground truth orbits. To compare the performance of our LSNN on symplectic integrations, we additionally employ the SWIFT\_RMVS3 symplectic integrator \citep{Levi1994} to generate the learning data for the same case of the 2:3 resonant KBOs studied in Section 4.2. However, when we let the LSNN to learn the symplectic ground truth orbits, the resulting predictions are not good, as shown in Fig. \ref{SymInt}. We can observe that, for the validation particle tp1 staring with $\sigma_0=60^\circ$, the predicted orbital parameters (red dots) deviate significantly from the ground truth data (black dots). The absolute error in the Jacobi integral (left panel) can reach a value as large as 0.031, which is over three times larger than that displayed in Fig. \ref{stability}. It is well known that in the PCR3BP, the Tisserand relation gives an approximate relationship between the Jacobi integral $C$ and the semimajor axis $a$ of the particle, as
\begin{equation}\label{tisserand}
C \approx \frac{1}{a}+2\sqrt{a(1-e^2)}, 
\end{equation}
where the eccentricity $e$ was set to have a very small value of 0.01. Therefore, the large error in $C$ will induce a considerable drift in $a$, as seen in the right panel of Fig. \ref{SymInt}.

The failure of our LSNN when using the symplectic integrator to generate the learning data may be due to its lower accuracy compared to the RK integrator. This is particularly evident in the ground truth Jacobi integral, for which the numerical error of the RK integrator is as small as $<10^{-12}$, while the symplectic integrator only provides a solution that is precise to $10^{-3}$. \citet{CT21} also noticed this issue, and they suggested that the symplectic integrator is better suited for separable Hamiltonian systems (e.g. the 2-body problem), while the non-symplectic integrator is preferable for non-separable ones (e.g. the 3-body problem). Furthermore, in Tao's earlier work \citep{T16}, the author specifically pointed out that, for the long time numerical simulations, the non-symplectic integrator is more accurate and efficient than the symplectic integrator. This argument demonstrates that the RK method could be a great way to generate training data for machine learning.


\section*{Acknowledgments}

This work was supported by the National Natural Science Foundation of China (Nos. 11973027, 11933001, 11601159 and 12150009), and National Key R\&D Program of China (2019YFA0706601). The authors are grateful to the two referees for their valuable comments, which helped to considerably
improve this paper.

\section*{Data Availability}

The data underlying this article are available in the article and in its online supplementary material.

\end{document}